\documentclass[aps]{revtex4}
\usepackage{graphicx}
\begin{document}

\title{ Mergers
 of Irrotational Neutron Star Binaries in Conformally Flat Gravity}
\author{Joshua A.\ Faber}
\affiliation{Department of Physics and Astronomy, 
 Northwestern University, Evanston, IL 60208}
\author{Philippe Grandcl\'ement}
\altaffiliation[Current address: ]{Laboratoire de Math\'ematiques et de Physique
 Th\'eorique, Universit\'e de Tours, Parc de Grandmont, 37200 Tours, France} 
\affiliation{Department of Physics and Astronomy, 
 Northwestern University, Evanston, IL 60208}
\author{Frederic A.\ Rasio}
\affiliation{Department of Physics and Astronomy, 
 Northwestern University, Evanston, IL 60208}

\date{\today}

\begin{abstract}
We present the first results from our new general relativistic,
Lagrangian hydrodynamics
code, which treats gravity in the conformally flat (CF) limit.  The
evolution of fluid configurations is described using smoothed particle
hydrodynamics (SPH), and the
elliptic field equations of the CF formalism are solved using spectral methods
in spherical coordinates.  The code was tested on models for which the
CF limit is exact, finding good agreement with the classical
Oppenheimer-Volkov solution for a relativistic static spherical star
as well as the exact semi-analytic solution for a collapsing
spherical dust cloud.  By computing the evolution of quasi-equilibrium
neutron star binary configurations in the absence of gravitational radiation
backreaction, we have confirmed that these configurations can remain
dynamically stable all the way to
the development of a cusp.  With an approximate treatment of
radiation reaction, we have calculated the complete merger of an
irrotational binary
configuration from the innermost point on an equilibrium sequence
through merger and remnant formation and ringdown, finding good agreement with
previous relativistic calculations.  In particular, we find that mass
loss is highly suppressed by relativistic effects, but that, for a
reasonably stiff neutron star equation of state, the
remnant is initially stable against gravitational collapse because of
its strong differential rotation.  The gravity wave signal derived
from our numerical calculation has an
energy spectrum which matches extremely well with estimates based
solely on quasi-equilibrium results, deviating from the Newtonian
power-law form at frequencies below 1 kHz, i.e., within the reach of
advanced interferometric detectors.
\end{abstract}

\maketitle

\section{Introduction}

Gravitational wave (GW) astronomy stands at a crucial moment in its
history, with the LIGO (Laser Interferometer Gravitational Wave
Observatory) Scientific Collaboration reporting results from their first scientific
runs \cite{LIGO_S1,LIGO,LIGOweb}, GEO600 having completed two scientific runs
\cite{GEO,GEOweb}, TAMA taking data \cite{TAMA,TAMAweb}, and
VIRGO reporting its first lock acquisition \cite{VIRGO,VIRGO2,VIRGOweb}.
As such, it is now more important than ever
to have accurate theoretical predictions of the main candidate
signals, both to aid in their detection and to facilitate the
interpretation of any future detections.

It has long been recognized that coalescing relativistic binary
systems containing compact objects, either neutron stars (NS) or black
holes (BH), are likely to be important sources
of detectable GWs.  Recent population synthesis calculations
indicate that an Advanced LIGO detector should be
able to see at least tens of coalescences per year of NS-NS, NS-BH,
and BH-BH binaries \cite{BKB}.  Empirical rate estimates based on three of the
four observed binary pulsar systems expected to merge within a Hubble
time, PSR B1913+16, PSR B1534+12, and PSR J0737-3039 \cite{Burgay}, are in
general agreement, with the latter making the dominant contribution to
the probabilistic rate because of its short coalescence time of
$85~{\rm Myr}$
(see \cite{KKL,Chunglee} and references therein; the fourth 
system is located in
a globular cluster, rather than the galactic plane).
NS-NS binaries are the only known systems with coalescence times
shorter than a Hubble time to
have been conclusively observed, and it is the orbital decay of
PSR1913+16 that currently provides our best indirect evidence for the
existence of GWs \cite{HT,Tay2,Tay3}.  

At large separations, the dynamics of compact object binaries can be well
approximated by high-order post-Newtonian (PN) (see
\cite{Blan} and references therein) and other formalisms
\cite{DIS1,DIS2} which treat
the compact objects as point-masses, including the effects of
spin-orbit and spin-spin
angular momentum couplings for systems containing a BH \cite{Kid,Ap1,Ap2}.  
In general, these methods are more appropriate for describing a BH
than a NS, since the finite-size effects ignored by point-mass
formulae are of greater magnitude in systems containing NS.

For smaller binary separations, $r \lesssim  10 R_{NS}$, where $R_{NS}$
is the radius of the NS, these
finite-size corrections play an increasingly significant role in the
evolution of NS-NS binaries.  As long as the coalescence timescale
$\tau=r/\dot{r}$ remains longer than the dynamical timescale, the
evolution is quasi-adiabatic, and the binary will sweep through a
sequence of configurations representing energy minima for given
binary separations.  The infall rate will be given by 
\begin{equation}
\frac{dr}{dt}=\left(\frac{dE}{dt}\right)_{GW}\left(
\frac{dE(r)}{dr}\right)^{-1}_{eq},
\end{equation}
where $(dE/dt)_{GW}$ is the energy loss rate to gravitational
radiation, and $(dE/dr)_{eq}$ is the slope of the equilibrium energy curve.
Equilibrium energy sequences were first constructed in Newtonian
and then PN gravity (see \cite{RS99} and references therein). 
More recently, general relativistic (GR) 
sequences have been calculated for binary
NS systems in quasi-circular orbits
\cite{BGM2,GGTMB,TG,BCSST1,BCSST2,MMW2,UE,USE}.  
An important result from these studies is that the slope of the
equilibrium energy curve is flatter when relativistic effects are
included, compared to the Newtonian value.  This becomes especially
pronounced at
small separations, near the point where equilibrium sequences
encounter an energy minimum (often associated with the
``innermost stable circular orbit'', or ISCO) or terminate when a cusp develops
on the inner edge of the NS.  Based on this observation, it was noted
that a flatter slope in the equilibrium energy curve results in not
only a faster infall rate but also a decrease in the energy spectrum, 
$dE/df\approx (dE/dr)/(df/dr)$.  In fact, based on the relativistic
equilibrium sequences of Taniguchi and Gourgoulhon \cite[hereafter
  TG]{TG}, Faber {\it et al.} \cite[hereafter FGRT]{FGRT} concluded that the
dependence of finite-size corrections on the NS compactness
(i.e. $M_{NS}/R_{NS}$) would leave an imprint in the GW energy
spectrum at frequencies $f\approx 500-1000~{\rm Hz}$, within the band
accessible to Advanced LIGO.  Hughes \cite{Hughes} proposed a method using these
results which would allow for a determination of the NS compactness to
within a few percent based on $10-50$ observations with Advanced LIGO if it employs
a narrow-band detector in addition to the current broadband setup,
with the required number of observations dependent upon both the true
compactness and the particular setup of the detector.

Shortly before the ISCO or the end of the equilibrium sequence (when
it terminates at the formation of a cusp) the binary
will begin a transition toward a rapid plunge inward, eventually
leading to merger. 
Once the binary passes this point, the dynamical evolution
becomes too complicated to describe using semi-analytical methods,
requiring instead a 3-d hydrodynamic treatment.  Such calculations were
performed first for Newtonian gravitation, 
using both Eulerian, grid-based codes and
Lagrangian smoothed particle hydrodynamics (SPH; for a review, see,
e.g.,  \cite{FR3} and references therein).
It was always recognized that any results would
be at best qualitatively accurate, since the extreme
compactness of the NS would induce a host of GR effects.   
Noting this, some attempts were made to
calculate the evolution of binary NS systems in lowest-order PN
gravity \cite{FR1,FR2,FR3,SON3,APODR}, using a formalism developed by
\cite{BDS}.  Unfortunately, using realistic NS equations of state
(EOS) violated the basic assumption of the PN approximation that 
the magnitude of the 1PN terms are small relative to Newtonian-order effects.
As a result, all PN calculations were forced to make unphysical approximations,
either by evolving NS with a fraction of their proper physical mass
\cite{SON3,APODR}, or by reducing the magnitude of all 1PN terms
\cite[hereafter denoted FR1--3, or collectively FR]{FR1,FR2,FR3}.

While the ultimate goal of studying binary NS coalescences should be a
full GR treatment, only one group has been able to calculate the full
evolution of a binary system from an equilibrium binary configuration
through merger and the formation of a remnant \cite{SU1,SU2,STU}.
While these calculations represent a breakthrough in our understanding
of the hydrodynamics of coalescing NS binaries, they still leave a
great deal of room for further research into a variety of questions.
Currently, full GR calculations are extremely difficult, and are
vulnerable to several numerical instabilities.  In order to
guarantee the stability of calculation through coalescence and the
formation of either a stable merger remnant or a BH, binaries were 
started from the termination points of equilibrium sequences, from
quasi-circular configurations with zero infall velocity. The
calculations of Shibata and Uryu \cite{SU1,SU2} used initial
conditions generated by Uryu, Eriguchi, and collaborators
\cite{UE,USE}, and the more recent work by Shibata, Taniguchi, and
Uryu \cite{STU} used initial conditions generated by TG. 
This restricts studying the behavior of the infall velocity and GW signal when
the stars begin their plunge, during which time detected signals may yield
important information, as mentioned above.  
The lack of information about the dynamics of
binaries before the plunge also hinders testing the
validity of the initial matter and field configurations used in full
GR calculations.  While the assumptions defining the quasi-equilibrium
configuration are  
certainly reasonable, it is difficult to gauge 
how well such configurations agree with those evolved dynamically
from larger separations.  Also unknown in detail is the
effect that errors in the initial configuration will have on the
system as they propagate in non-linear fashion during the calculation.
 
A further technical difficulty with full GR calculations results from
computational limits, since numerical grids are currently
constrained to have their boundaries lie within approximately half of
a GW wavelength, within the near
zone.  This can induce possibly significant errors into the GW extraction
process, which would ideally be performed by studying the behavior of the
metric in the wave zone.

A possible middle road, at least at present, is provided by the
Conformally Flat (CF) approximation to GR, first described by
Isenberg \cite{Isen} and developed in greater detail by
Wilson and collaborators 
\cite{WMM} (note that their original version contained a mathematical
error, pointed out by Flanagan \cite{Flan}, accounting for spurious results
with regard to ``crushing'' effects on NS prior to merger).  This
formalism includes much of the non-linearity inherent in GR, but
yields a set of coupled, non-linear, elliptic field equations, which
can be evolved stably.  The first dynamical 3-d calculations to make
use of the CF framework were performed by Shibata, Baumgarte, and
Shapiro \cite{SBS}, who created a
PN variant by discarding some of the non-linear source terms in the
field equations while retaining the vast majority of the non-linearity
in the system.  They found, among other things, that the maximum
density of the NS is smaller for binary configurations than in
isolation, and that NS in binaries have a higher maximum mass as well,
strongly indicating that collapse to a BH prior to merger is
essentially impossible.  The first
calculations to use the full formalism were performed by Oechslin 
{\it et al.} \cite{Oech},
using a Lagrangian SPH code with a multigrid field solver.  
Using corotating initial configurations, which are thought to be
unphysical (since viscous effects are thought to be much too weak to 
tidally synchronize the NS \cite{BC,Koch}), they found that mass loss
during the merger is suppressed in relativistic gravitational schemes,
as had been previously suggested based on calculations in PN gravity
(FR3). 

While the CF formalism appears to be safer than full GR with regard to
evolving the system stably, the grid-based approaches used so far
suffer from many of the same problems faced by full GR calculations.
Since the CF field equations are non-linear and lack compact support,
approximate boundary conditions for the fields must be applied at the
boundary of the grid, which can lead to errors in the solution.
Additionally, very large grids are required to satisfy the tradeoff
between large grid sizes, required to improve the accuracy of the
boundary conditions, and small grid cell sizes, necessary to improve
the resolution of the NS.  Multi-grid techniques can be very valuable
for such cases, but grid refinement
techniques can also act as a source of numerical errors.  Noting this,
we have taken an entirely new approach to the dynamical evolution of
binary NS systems with SPH, which requires no use of rectangular 3-d
grids to solve for the metric fields.  Instead, we use the
LORENE numerical libraries, developed by the Meudon group, which are
freely available online at {\tt http://www.lorene.obspm.fr} (see
\cite[hereafter BGM]{BGM} and \cite{Grand}
for a thorough description of the numerical techniques
used in LORENE).  These routines, which use spectral methods and
iterative techniques to solve systems of coupled non-linear
multidimensional PDEs, have been used widely to study quasi-equilibrium
sequences of binary NS systems in Newtonian \cite{TGB,TG2} and CF gravity
(TG,\cite{GGTMB,TG3}), as well as a number of other fluid configurations.  
The CF quasi-equilibrium
solutions of Gourgoulhon {\it et al.} \cite[hereafter GGTMB]{GGTMB}
and TG
have been used as the initial configurations for the most recent
full GR calculations of Shibata, Taniguchi, and Uryu, 
\cite{STU}, and will be used in this work as well.

Our work here represents the first time that spherical coordinates and
spectral methods have
been used to study the dynamical evolution of binary NS systems in
any gravitational formalism, including Newtonian gravity.  It has long
been known that these techniques are ideal for describing both binary
systems with large separations
as well as the merger remnants formed during the coalescence, since
the spherical coordinates correspond much
more naturally to the metric fields than rectangular coordinates can.
Traditionally, rectangular grids have been used anyway, both because they
are more widespread throughout computational fluid dynamics, 
but also because they are viewed as more robust.  It has often been
assumed that spherical coordinate field solvers may fail
to calculate fields properly during the merger, when the matter can be
described neither as two spheroids nor as one, but rather some
combination of the two, with mass loss streams and other phenomena
confusing the picture.  We show here, however, that spectral
methods can be used successfully in this regime, taking advantage
of the 
multi-domain techniques of LORENE.  This  allows us to take advantage of
the many advantages inherent to spectral methods: improved speed,
vastly improved computer memory efficiency, and a coordinate system
which allows for a natural treatment of the exact boundary conditions.

The code we have developed to perform 3-d, relativistic, numerical
hydrodynamic evolutions is well-suited for the study of a number of
physical systems.  While we focus here on
merging binary NS, our code is capable of evolving essentially
any binary or single-body relativistic fluid configuration.  These
include collapsing stellar cores and supermassive stars,
and rapidly rotating fluid configurations.  Modules currently exist to
handle a number of physically-motivated EOS, including models with
phase transitions and bosonic condensates, and we plan to extend our studies to
include appropriate treatments of these conditions in the future.

Our paper is structured as follows.  In Sec.~\ref{sec:numerical}, we
summarize the theoretical basis for our new relativistic Lagrangian 
code, describing in turn
the numerical methods used to implement both
the dynamical equations of the  CF formalism in SPH and the details of 
our spectral methods field solver.  In Sec.~\ref{sec:calculate}, we
turn to the computational aspects of the code relevant to coalescing
binary systems.  We
detail the choices we have made with regard to SPH discretization, the
techniques used to convert between particle-based and spectrally
decomposed descriptions of various quantities, and the coordinate
transformations implemented 
to describe binary systems, merging systems, and the resulting remnants.
In Sec.~\ref{sec:tests}, we report the results of several test
calculations, including two well-known exact CF solutions as well as 
the evolution of quasi-equilibrium configurations in the absence of 
dissipative effects, producing circular orbits.
In Sec.~\ref{sec:results}, we show our results from the calculation of
a full NS binary coalescence started from the innermost
quasi-equilibrium configuration, and followed through
the merger and the formation of a merger remnant. We compare our
results to previous work in the field, including full GR calculations
using the same initial condition.  Finally, in
Sec.~\ref{sec:final}, we summarize our results and lay out some of the
many classes of problems in relativistic hydrodynamics where our code
may prove useful.

\section{Numerical Methods}\label{sec:numerical}

The CF approximation assumes that the spatial part of the GR
metric is equal to the flat-space form, multiplied by a
conformal factor which varies with space and time.  Setting $G=c=1$,
as we will do throughout this paper, the CF metric takes the
form
\begin{equation}
ds^2=-(N^2-N_i N^i)dt^2-2N_i dt dx^i+A^2 f_{ij}dx^i dx^j,
\label{eq:metric}
\end{equation}
where $N$ is the lapse function, $N_i$ is the shift vector, and $A$
will be referred to here as the conformal factor (see Table \ref{table:not}
for a comparison of our notation to those of
\cite{GGTMB,Oech,SBS,WMM}, which are all
based on the same exact assumptions).  The flat-space three-metric is
denoted by $f_{ij}$.  We follow the standard notation
for relativistic tensors, denoting spatial three-vectors with Latin
indices, and relativistic four-vectors with Greek indices.
While the CF approximation cannot reproduce the exact GR solution for an
arbitrary matter configuration, 
it is exact for spherically symmetric systems, and yields field solutions
which agree with those calculated using
full GR to within a few percent for many systems of
interest \cite{MMW}.

As is typical for any relativistic calculation utilizing a polytropic
EOS, there is a single physical scale which defines the units
for the problem.  We choose to define this scale for single-body test
calculations (Secs. \ref{sec:ov} and \ref{sec:dust}) by dividing all
mass, length, and time-based quantities by the gravitational (ADM)
mass $M_0$
of the object.  For all binary calculations, we scale our results by the
``chirp mass'' of the system, ${\cal M}_{ch}\equiv \mu^{3/5} M_t^{2/5}$, where
$M_t$ is the total gravitational mass of the system at large
separation, and $\mu\equiv M_1 M_2/M_t$ is the reduced
(gravitational) mass.  This quantity is expected to be the most
directly measurable physical parameter deduced from any GW
observation (see, e.g., \cite{CF}).  For a binary system consisting
of two $M_0=1.4~M_\odot$ NS, the chirp mass is ${\cal
  M}_{ch}=1.22~M_\odot$, which yields characteristic distance and time scales
$R=1.80~{\rm km}$ and $T=6.00\times 10^{-6}~{\rm sec}$,
respectively.  To compare our results with the equivalent relativistic
model (run M1414) of \cite{STU}, who
use the total gravitational mass of the binary at large separation as
their unit, one can divide our quoted masses and radii 
by a factor of $2^{1.2}=2.30$.  To convert our time evolution results
into the time units used in \cite{STU}, which are
defined in terms of the initial binary orbital period, divide our time
units by a factor of $443$ (their unit $P_{t=0}=2.66~{\rm ms}$ for two
NS each with $M_0=1.4~M_\odot$). 

\subsection{CF Smoothed Particle Hydrodynamics}

In what follows,
we will assume that the stress-energy tensor of the fluid is that
of a perfect fluid, with
\begin{equation}
T_{\mu\nu}=(\rho+\rho\epsilon+P)u_{\mu}u_{\nu}+Pg_{\mu\nu},
\label{eq:stressenergy}
\end{equation}
where $\rho$ is the rest-mass density, $\epsilon$ is the specific
internal energy, $P$ the fluid pressure, and $u_{\mu}$ the
four-velocity of the fluid.  For our initial data, we assume a
polytropic EOS $P=k\rho^\Gamma$, with constant values
for $k$ and the adiabatic index $\Gamma$, and throughout the
calculation we assume that the evolution is adiabatic, such that
$P=(\Gamma-1)\rho\epsilon$.  The maximum infall velocity of the two
stars relative to each other during our dynamical calculations was found to be
$v_{in}=0.06{\rm c}$, whereas the sound speed at the center of the NS
is initially $c_s=0.85{\rm c}$.  Since the sound speed for our
EOS depends on density such that $c_s\propto \rho^{1/2}$,
we expect that the only supersonic fluid flows will occur in the very
tenuous outer regions of the NS.  During the merger, we do expect that
low-density matter from the inner edge of each
NS will cross over to
the binary companion at high relative velocity ($\delta v \approx
0.3{\rm c}$), but the velocity field is dominated by the circular
motion, rather than a converging flow.   All motion within the cores of
the NS should remain strongly subsonic throughout the merger process.

Using our CF metric and stress-energy tensor, Eqs.~\ref{eq:metric} and
\ref{eq:stressenergy}, 
the Lagrangian continuity equation is given by
\begin{equation}
\frac{d\rho_*}{dt}+\rho_*\nabla_i v^i=0,
\label{eq:continuity}
\end{equation}
where the rest mass density is defined by 
\begin{equation}
\rho_*\equiv Nu^0 A^3 \rho=\gamma_n A^3\rho,
\label{eq:rhostar}
\end{equation}
the Lorentz factor of the fluid is defined as
\begin{equation}
\gamma_n\equiv Nu^0,
\label{eq:lorentz}
\end{equation}
and the physical velocity is given by
\begin{equation}
v^i\equiv u^i/u^0=N^i+\frac{u_i}{A^2u^0}.
\label{eq:uv}
\end{equation}
Note that $u^0$ is the timelike element of the covariant 4-velocity;
all other numerical superscripts refer to exponents.
Following the SPH prescription, this conservative form of the continuity
equation allows us to define
a set of particles, each of which has a fixed mass $m_a$ and a well
defined velocity given by $dx_i/dt=v_i$.  
For each particle ``a'', we also define a ``smoothing length'', $h_a$,
which represents the physical size of the particle.  SPH particles do
not have delta-function density profiles; rather, each particle
represents a spherically symmetric density distribution centered at
the particle's position $\vec{x}_a$ with compact support.  This
density distribution, determined by our chosen form of the smoothing
kernel function, is second-order differentiable, and drops to zero at a
radius equal to two smoothing lengths from the particle's position.
For each particle, we define a set of neighboring particles by the
condition that all particles whose centers
fall within a given particle's compactly supported
density distribution are its neighbors.  To determine the proper
smoothing length for each particle, we define an ideal number of
neighbors for each particle, $N_N$, and we use a relaxation technique
to adjust $h_a$ after every timestep (as described
in detail in, e.g., FR1 and \cite{RS1}).  We note in passing that
``neighborhood'' is not a reflexive property; we handle this through
the use of a ``gather-scatter'' algorithm (see \cite{StarCrash} for details).

The primary advantage of the SPH method over traditional grid-based
codes is that fluid advection is handled in a natural way, such that
one can define the edge of a fluid distribution without recourse to
artificial ``atmospheres'' or other tricks necessary to prevent matter from
bleeding into the vacuum.  Particles simply follow their trajectories
in the fluid flow.  As such, SPH is extremely computationally
efficient, since all numerical resources are focused automatically on those
regions containing matter.  Because of this, SPH also allows
for high spatial resolution.  The primary disadvantage of SPH
compared to shock-capturing grid-based codes is in the lower resolution of shock fronts,
which, as we discuss below, is unimportant here.

All hydrodynamic quantities
can be defined using standard SPH summation techniques over each
particle's neighbor list (see,
e.g., \cite{RS1,Lom2,Lom3} for a thorough discussion), with the rest
mass density $\rho_*$ taking the place of the standard Newtonian density.
Thus the rest-mass density for particle ``a'' can be defined via SPH
summation over a set of neighboring particles denoted by ``b'' as
\begin{equation}
(\rho_*)_a=\sum_b m_b W_{ab},
\label{eq:dens_sph}
\end{equation}
where $W_{ab}$ is the $W_4$ smoothing kernel
function for a pair of particles first introduced by Monaghan and
Lattanzio \cite{ML}, used
in FR, \cite{RS1} and many other implementations.
The momentum equation is given by 
\begin{equation}
\frac{d\tilde{u}_i}{dt}=-\frac{NA^3}{\rho_*}\nabla_i P - 
Nhu^0\nabla_i N-\tilde{u}_j\nabla_i
N^j+\frac{\tilde{u}_k\tilde{u}_k}{A^3hu^0}\nabla_i A,
\label{eq:dutildedt}
\end{equation}
where the specific momentum is defined by 
\begin{equation}
\tilde{u}_i\equiv hu_i,
\label{eq:utilde}
\end{equation}
and the specific enthalpy is defined as 
\begin{equation}
h\equiv(1+\epsilon+P/\rho)=1+\Gamma\epsilon.
\label{eq:enthalpy}
\end{equation}
In this expression and throughout this paper,
covariant derivatives are associated with the flat-space metric.
In the absence of non-adiabatic
artificial viscosity terms,
the energy equation merely implies that the value of $k$ in the EOS
remains constant.
In our calculation, we use $\rho_*$ and $\tilde{u}_i$ as the basic
hydrodynamic variables (in addition to our uniform value of $k$).  To
find $u^0$, which enters into the momentum equation, we take the
normalization condition for the 4-velocity, $u_{\mu}u^{\mu}=-1$, and
find
\begin{equation}\label{eq:gamman}
{\gamma_n}^2=(Nu^0)^2=1+\frac{u_i
u_i}{A^2}=1+\frac{\tilde{u}_i\tilde{u}_i}{A^2}\left[1+\frac{\Gamma k
\rho_*^{\Gamma-1}}{(\gamma_n A^3)^{\Gamma-1}}\right]^{-2},
\end{equation}
which can be solved implicitly in terms of the density and the field values.

To solve the field equations of GR, we need to fix the slicing
condition for timelike hypersurfaces.  Following the standard approach
to the CF formalism,
we find that the extrinsic curvature tensor
is given in terms of the shift vector as
\begin{equation}
K^{ij}\equiv \frac{1}{2NA^2}\left(\nabla^i N^j +\nabla^j N^i
-\frac{2}{3} f^{ij}\nabla_k N^k\right),
\end{equation}
when we assume the maximal slicing condition (${\rm Tr} K=0$).    
To lower the indices of the
extrinsic curvature tensor and other spatially-defined quantities, we
use the conformal flat-space metric (i.e., $K_{ij}=A^4
\delta_{ik}\delta_{jl}K^{kl}$  in Cartesian coordinates, where
$\delta_{ij}$ is the Cartesian flat-space metric).  Combining
the maximal slicing assumption with the Hamiltonian constraint yields
a pair of elliptic equations for $\nu\equiv \ln(N)$ and $\beta\equiv
\ln(AN)$ (GGTMB), in the form
\begin{eqnarray}
\nabla_k\nabla^k \nu &=& 4\pi G A^2 (E+S) +A^2 K_{ij}K^{ij} -\nabla_i
\nu \nabla^i \beta, 
\label{eq:pois_nu}\\
\nabla_k\nabla^k \beta &=& 4\pi G A^2S +\frac{3}{4} A^2 K_{ij}K^{ij}
-\frac{1}{2}(\nabla_i\nu \nabla^i \nu +\nabla_i\beta\nabla^i\beta),
\label{eq:pois_beta}
\end{eqnarray}
where the matter energy density and the trace of the stress energy
tensor are given by
\begin{eqnarray}
E&\equiv& {\gamma_n}^2(\rho h)-P, \label{eq:defn_E}\\
S&\equiv& \frac{{\gamma_n}^2-1}{{\gamma_n}^2}(E+P)+3P. \label{eq:defn_S}
\end{eqnarray}
Note that Eqs.~\ref{eq:pois_nu} and \ref{eq:pois_beta}
are algebraically equivalent to the field equations found in other
papers on the CF formalism, although most groups have typically solved
the corresponding Poisson-type equations for $\psi\equiv\sqrt{A}$ and $N\psi$.
Finally, the momentum constraint gives us the equation for the shift
vector, 
\begin{equation}
\nabla^j\nabla_j N^i+\frac{1}{3}\nabla^i\nabla_j N^j = -16\pi G
 \frac{N}{h\gamma_n}(E+P)\tilde{u}_i+2NA^2K^{ij}\nabla_j(3\beta-4\nu).
\label{eq:pois_shift}
\end{equation}
Since the CF formalism is time-symmetric, and we can
define several conserved quantities.  The total baryonic mass,
\begin{equation}
M_b=\int \rho_* d^3 x,
\label{eq:restmass}
\end{equation}
is automatically conserved in our SPH scheme, since we use the rest
mass density to define particle masses.  The total angular momentum of
the system can be defined as
\begin{equation}
J_i=\varepsilon_{ijk}\int \rho_* x^j\tilde{u}_k d^3x.
\label{eq:angmom}
\end{equation}
Finally, the ADM mass is given by
\begin{equation}
M_0=\int \rho_{ADM} d^3 x;~~~~
\rho_{ADM}\equiv A^{5/2}\left(E+\frac{1}{16\pi G}K_{ij}
K^{ij}\right),
\label{eq:admmass}
\end{equation}
It is important for numerical reasons to note that the two terms that
make up the ADM mass density $\rho_{ADM}$ have different behaviors:
the contribution from the matter energy density,
\begin{equation}
\rho_1\equiv A^{5/2}E,
\label{eq:rho1}
\end{equation}
has compact support, and is non-zero
only in the presence of matter, whereas the term involving the
extrinsic curvature,
\begin{equation}
\rho_2\equiv \frac{1}{16\pi G}A^{5/2}K^{ij}K_{ij},
\label{eq:rho2}
\end{equation}
extends throughout all space with power-law fall-off at large radii.

\subsection{The Spectral Methods Field Solver}

The spectral methods techniques we use
to solve the CF field equations, Eqs.~\ref{eq:pois_nu},
\ref{eq:pois_beta} and \ref{eq:pois_shift}, discussed in great detail
in \cite{Grand},
provide a number of important advantages not present in
traditional grid-based approaches.  First and
foremost is their speed and computational efficiency.  
Finite differencing schemes typically require 3-d grid sizes of
$\gtrsim 10^6$ elements \cite{Oech,STU}.  In contrast, GGTMB show that 
spectral method field solvers can be used to
construct field solutions yielding ADM masses and angular momenta 
convergent to within $10^{-4}$ and satisfying the virial theorem to the
same level using only 3 grids of size $17\times
13\times 12$.  These grids, extremely small compared to
those used in Cartesian multi-grid solvers,
result in a great increase in speed.  Additionally, the use of
spherical coordinates
allows for a more natural
treatment of boundary conditions.  In the approach
we use, taken from GGTMB
and summarized here, space is
described using spherical coordinates, split into a set of
nested ``computational domains''.  The outermost domain can be
compactified by rewriting the field equations in terms of $1/r$,
allowing us to impose
the exact boundary conditions at infinity,
rather than ``fall-off'' boundary conditions which
approximate the behavior of the fields at the edges of a rectangular
grid.  As such, we avoid the classic tradeoff between a grid with
large grid spacing, which yields accurate boundary conditions but poor
spatial resolution of the matter source, and a grid with small
spacing, and the opposite concerns.  

\begin{figure}
\centering\includegraphics[width=6in]{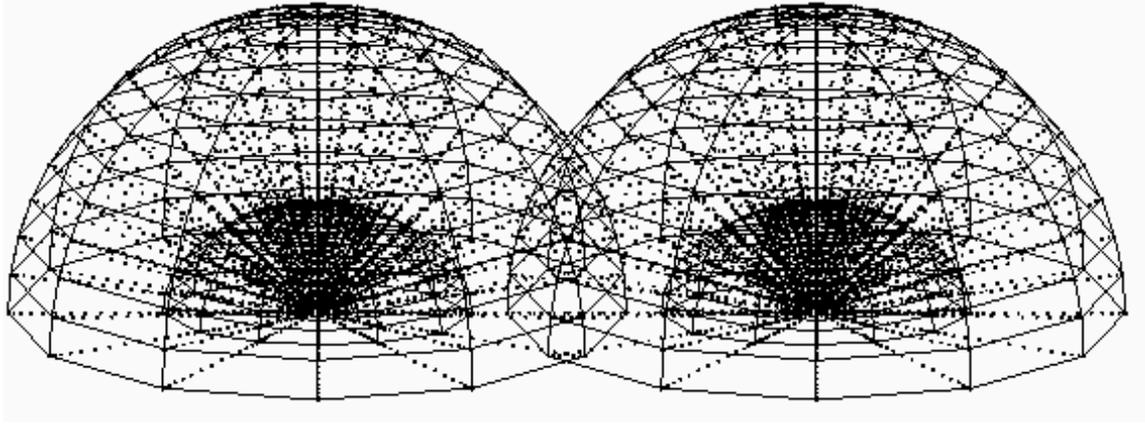}
\caption{Radial domains used to evaluate the field equations of the CF
method.  The boundaries of the inner two domains are shown as
lattices, with all collocation points within these domains shown as
well.  The outermost domain, 
which extends to spatial infinity, is not shown.}\label{fig:domains}
\end{figure}

Combining the LORENE methods with particle-based SPH requires some
small but significant changes from the previous approach described in
detail in Sec.~IVA of GGTMB.  As in that work, we construct a set
of three computational domains around each star to evaluate the field
equations.  The innermost domain has a spheroidal topology, with a
boundary roughly corresponding to the SPH particle configuration's
surface, as described
in Sec.~\ref{sec:surface}.  The other two domains cover successively
larger regions in radii, with the outermost domain extending out to
spatial infinity, as shown in Fig.~\ref{fig:domains}.  The field
equations are solved in each domain and the global solution is
obtained by matching the function and its first radial derivative at
each boundary.  Appropriate boundary conditions at radial infinity are
also imposed.  All fields can be described in one of two
complementary representations, either in terms of their coefficients
in the spectral
decomposition, or by means of their values at a set of 
``collocation points''.  The coordinate
representations for these points are defined such that
the origin of the system describing
each NS is located at the position of the star's maximum density, 
The collocation points are
spaced equally in both $\sin\theta$ and $\phi$, as required by the
angular component of the spectral
decomposition.  The radii of the points are determined from
the collocation points of the Chebyshev polynomial expansion used for
the radial coordinate, as described in BGM and GGTMB.  
For all calculations described in this
paper, these domains consisted of a $17\times 13\times 12$ grid, in
terms of radial, latitudinal, and longitudinal directions,
respectively, an acceptable trade-off between speed and accuracy,
based on the description above.

A key feature of the LORENE libraries is their handling of binary
systems in a straightforward and natural way.  This involves
``splitting'' the source terms of the field equations into two
distinct components, each of which is centered on one of the stars in
the binary. 
Since the field equations in this case are non-linear
the split cannot be performed uniquely; 
the fields present around one star cannot
be determined from its source terms only.  Rather, this method seeks
to minimize the contribution of one star to the fields around the
other.   In practice, both field variables and hydrodynamic source
terms can be broken down, as shown in Eqs.~80--88 of GGTMB, such that
\begin{equation}
\nu=\nu_{<1>}+\nu_{<2>}=\nu_{<1>}+\nu_{<2\rightarrow 1>}=
\nu_{<1\rightarrow 2>}+\nu_{<2>}, {\rm ~~etc.}
\end{equation}
where quantities labeled $<1>$ and $<2\rightarrow 1>$ are defined at
the collocation points of star 1, and $<2>$ and $<1\rightarrow 2>$ at
those of star 2.  The autopotentials of each star, 
$\nu_{<1>}$ and $\nu_{<2>}$, are
primarily generated by matter from the star itself, while the
``comp-potentials'', $\nu_{<2\rightarrow 1>}$ and $\nu_{<1\rightarrow
2>}$ are primarily generated by the other star.  It is this conversion
of fields between the two sets of coordinates which represents the
greatest amount of numerical effort during a calculation.  In
practice, we attempt 
to minimize the magnitude of the comp-potentials since they are
centered around the other star and not as well described by spherical
coordinates.  The detailed
description of how these quantities are defined and calculated can be
found in Sec.~IVC of GGTMB.

\section{Numerical Techniques for Coalescing Binaries}
\label{sec:calculate}

Integrating the LORENE library routines into an SPH-based
Lagrangian code introduces a number of rather subtle
numerical issues.  The simplest of these is deciding the shape of the
innermost computational domain's surface.  
The closer the boundary of the innermost domain lies to the
surface of the fluid, the more Gibbs effect errors are minimized, so
long as the surface is sufficiently smooth and convex everywhere.  
Gibbs effects, which are common to all spectral decomposition
techniques, result from the attempt to describe a non-differentiable
density distribution as a weighted sum of smooth functions; they
are relevant here when we attempt a spectral decomposition of a region
of space where the density drops to zero within the boundary (see BGM
for more details).
Unfortunately, the smoothness and convexity conditions do not necessarily apply
to the region in space where the SPH density is non-zero, since a
single SPH particle being shed from the star can greatly affect the
non-zero density boundary, even though it may represent an insignificant amount
of mass.  As a result, we have implemented an algorithm that attempts
to take the middle ground, defining a boundary for the innermost
domain which encloses as much of the mass as
possible, in such a way that smoothness and convexity are guaranteed.
The second domain is bounded by a sphere of radius twice that of the
outermost point of the first domain, and a third domain extends to
spatial infinity.

Once the configuration of the computational domains is determined,
there are several choices which need to be made with regard to the
most accurate and efficient way to calculate various terms in the
evolution equations.  Some
hydrodynamic quantities, such as the rest mass density $\rho_*$, need to be
defined for each of the SPH particles.  Here, we use $N\sim 100,000$
SPH particles for each run, which was found to be sufficient for
achieving numerical convergence of the GW signal to the $\sim 1-2\%$
level in our studies with post-Newtonian SPH \cite{FR3}.
Other quantities, such as those appearing in the source terms for the
field equations, need to be defined at every point among the $17\times
13 \times 12$ spheroidal grids of collocation points for each of the
three domains.  Compared to solutions computed using larger grid
sizes, we find that these agree with significantly larger grids to
within $\sim 0.1\%$ for the value of the shift vector, and even better
for the values of the lapse function and conformal factor.

In general, however, many quantities do not need to
be defined both ways, so long as a full set of thermodynamic
variables is known in both representations.  Details about which
quantities are used in each representation are given below.
Briefly, it is most efficient to perform
the majority of our algebraic operations on quantities defined at collocation
points, reading in and exporting back as small a set of parameters as
possible to the full set of SPH particles.  While reading quantities
from SPH particles to collocation points is relatively quick,
requiring an SPH summation at every collocation point position, the
reverse process is much more involved.  To calculate the value of a
quantity known in the spectral representation at SPH particle
positions requires performing a sum over all spectral
coefficients with the weights appropriate to each particle
position, consuming a great deal of time. 

\subsection{Binary systems}
\label{sec:surface}
To construct the initial SPH particle configuration for a binary NS evolution,
we use the quasi-equilibrium irrotational models of TG, which are
publicly available at {\tt http://www.lorene.obspm.fr/data/}.  
These models describe the
complete 3-dimensional structure of both the field values and
hydrodynamic quantities at every point in space, in terms of a
spectral decomposition.
Specifically, we take the
results of their irrotational run for stars of equal mass and
equal compactness $GM/Rc^2=0.14$ with a $\Gamma=2$ polytropic
EOS. Each star has a baryonic mass in isolation given by
$GM_B/Rc^2=0.1461$.  

To convert
these models, which are stored in the coefficient basis, into
spatially defined particle-based quantities, we first lay down a grid
of SPH particles in a hexagonal close-packed (HCP) lattice with constant
lattice spacing and particle smoothing length.  This grid is then
treated as if we reflected around 
the $z=0$ plane, to take advantage of the vertical
symmetry inherent in the problem.   Each particle is treated as if it
were really two particles for all SPH summations, 
one located above the $z=0$ plane, the
other an equal distance below, both with half the true mass of the
``real'' particle.  Since the vertical symmetry is
enforced on a particle-based level, we solve all our field equations
only for vertical angles $0<\theta<\pi/2$, and reflect the solutions
for all points below the plane.
The mass of each particle is initially set to be
proportional to the density at the particle's position
according to the quasi-equilibrium model.  

Next, using an iterative process, we calculate the SPH expression
for the density of each particle, using Eq.~\ref{eq:dens_sph}, 
and adjust each particle's mass so
that the SPH density matches the proper value from the initial model,
stopping once the maximum difference for any particle is less than $0.25\%$ of
the star's central density.  Particle velocities
are assigned to match the quasi-equilibrium model's velocity field,
as are all other thermodynamic variables.  Finally, we advance the velocities
by half a time-step, using the same methods as in a standard iteration
loop (see below), since we use a second-order accurate
leapfrog algorithm (described in detail in \cite{RS1}).

During each iteration, the first step is always the calculation of
each particle's neighbor list, and the associated SPH forms for the
density and other hydrodynamic terms.  Once this is done, we perform a
Euclidean transformation on our coordinates into a new frame (denoted
by primed quantities), whose origin is defined to be the system center-of-mass, with
the center-of-mass of each star lying on the x-axis, making sure to transform the positions,
velocities, and accelerations for each particle.  In terms of the
inertial frame coordinates of the NS centers-of-mass,
$\vec{x}_1$ and $\vec{x}_2$, the transformation is given by
\begin{eqnarray}
\phi'&\equiv&\tan^{-1} \frac{y_2-y_1}{x_2-x_1}, \label{eq:phiprime}\\
x_{CoM}&\equiv&\frac{x_1+x_2}{2};~~ y_{CoM}\equiv\frac{y_1+y_2}{2},\\
x'(x,y)&\equiv& (x-x_{CoM})\cos\phi'+(y-y_{CoM})\sin\phi',\\
y'(x,y)&\equiv& (y-y_{CoM})\cos\phi'-(x-x_{CoM})\sin\phi',\\
\vec{x}'_{1,2}&=&(\pm \frac{R}{2},0,0); R\equiv\sqrt{(x_2-x_1)^2+(y_2-y_1)^2},
\label{eq:xprime12}
\end{eqnarray}

\begin{figure}
\centering\includegraphics[width=5.1in]{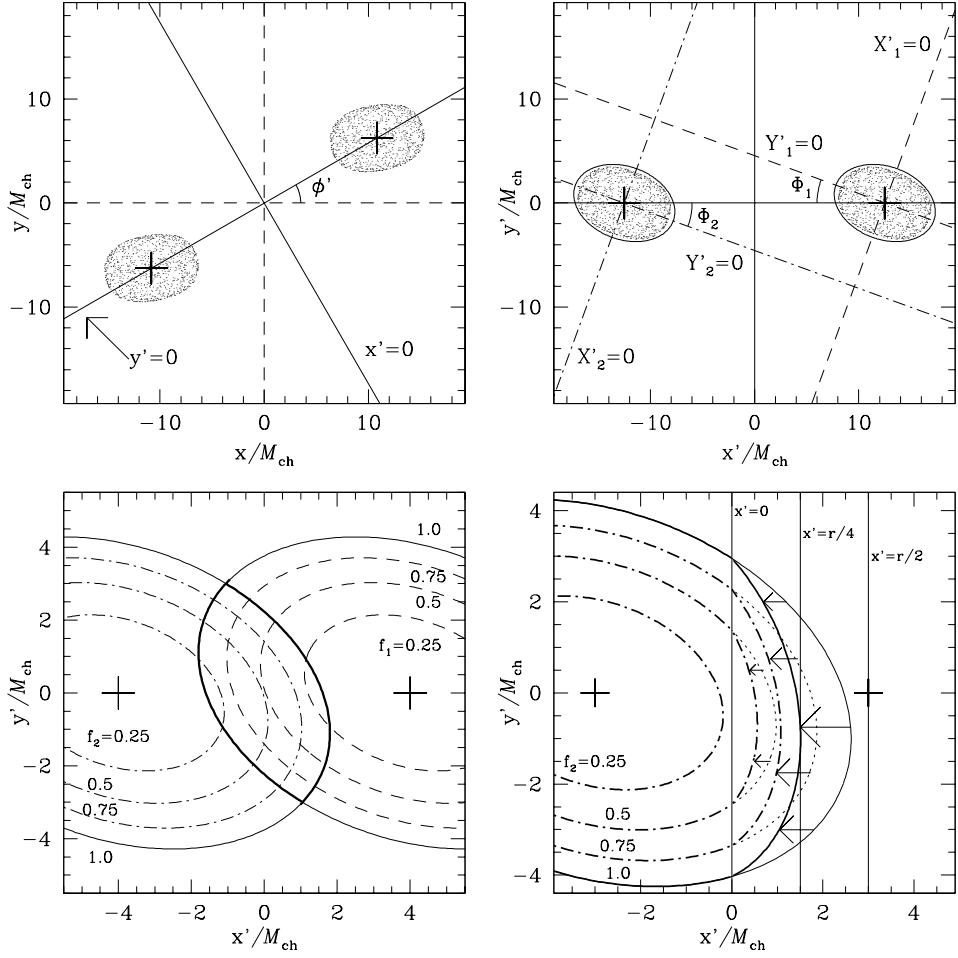}
\caption{A pictorial demonstration of the coordinate transformations
described in Secs.~\protect\ref{sec:surface} and
\protect\ref{sec:merging}; the particle configuration is for
demonstration purposes only, and was not taken from our calculations.  
In the upper left panel, we show a NS-NS
binary in the $x-y$ (inertial) frame, as well as the $x'-y'$ frame,
defined so that the centers-of-mass of the stars (crosses) lie
equidistant from the origin on the $x'$-axis (see Eqs.~\protect\ref{eq:phiprime}--\protect\ref{eq:xprime12}).  In the upper right, we
see the same system in the $x'-y'$ frame.  The angles $\Phi_1$ and
$\Phi_2$ are determined from the respective moment of inertia
tensors using Eq.~\protect\ref{eq:phiq}.  The best fit ellipsoidal
configurations, determined from
Eqs.~\protect\ref{eq:rthetaphi}--\protect\ref{eq:f0} are shown around
each star, aligning with the $X'_q-Y'_q$ axes, determined from
Eqs.~\protect\ref{eq:bigxprime} and \protect\ref{eq:bigyprime}.
In the bottom left, we show isocontours for the radial functions $f_1$ and
$f_2$, defined by Eq.~\protect\ref{eq:fq}, as well as the boundary of
the overlap region (heavy solid line).  Finally, in the
bottom right, we show the rescaling of the surface function for
very close configurations, showing only star 2 for clarity.  Here, the
binary separation is $r/{\cal M}_{ch}=6.0$, implying the maximum extent of the
surface of star 2 is to $x'/{\cal M}_{ch}=r/4{\cal M}_{ch}=1.5$.  
We linearly rescale the surface
function, as well as the corresponding values of $f_2$, for all points
with $x'>0$.}
\label{fig:coords}
\end{figure}

This transformation is shown in the upper left panel of Fig.~\ref{fig:coords}.
The next computational task is defining the shape of the innermost
computational domains,
calculated for sets of rays equally spaced in $\theta$ and $\phi$, as
measured from the center-of-mass of each star.  
We denote these surface functions $r(\theta_q,\phi_q)$, where the ``q''
subscript, taking the value $1$ or $2$, refers to angles measured in
the primed frame of Eqs.~\ref{eq:phiprime}--\ref{eq:xprime12} outward
from the center-of-mass of star $q$.  
These surfaces are used
to determine the position of the collocation points, which in turn are used as
the basis for the entire spectral decomposition.
While it is easy to
find the point along any ray
where the SPH density drops to zero, $r_{SPH}(\theta_q,\phi_q)$, we
have found that such a set of points leads to unacceptable results
from the field solver, especially with regard to the convexity of the
matter distribution.  If we ignore the density contributions of all
particles whose density falls below some fixed value,
say $\rho_{min}=0.0001$, the resulting
surface function $\tilde{r}_{SPH}(\theta_q,\phi_q)$ is typically much more
regular, but still is not an ideal choice, since we still cannot
guarantee convexity.
It should be noted that the field solver is entirely capable of
handling matter sources which 
lie outside the innermost domain, although it does work best in the
case where the surface of the matter matches the domain boundaries closely.
For this reason, we
restrict the shapes of the innermost domains to  triaxial ellipsoidal
configurations, oriented along the principal axes of the moment of
inertia tensor for each star.  The growing misalignment between the
stars and the (rotated) coordinate system is a well-known effect,
reflecting the tidal lag that develops in close binaries as matter
tries to reconfigure itself in response to the rapidly changing
gravitational field /cite{LRS5}.
Thus, for each star, we calculate the angle $\Phi_q$ such that
\begin{equation}
\Phi_q=\frac{1}{2}\tan^{-1}\left(\frac{2I_{xy}}{I_{xx}-I_{yy}}\right),
\label{eq:phiq}
\end{equation}
where $I_{ij}\equiv \sum_a m_a (x'_a-x'_q)_i (x'_a-x'_q)_j$, and
define our surface function $r(\theta_q,\phi_q)$ such that 
\begin{equation}
r(\theta_q,\phi_q)=\left[\frac{\hat{x}^2}{a^2}+\frac{\hat{y}^2}{b^2}+
\frac{\hat{z}^2}{c^2}\right]^{-1/2},\label{eq:rthetaphi}
\end{equation}
where 
\begin{eqnarray}
\hat{x}(\theta_q,\phi_q)&\equiv&\sin\theta_q
(\cos\Phi_q\cos\phi_q+\sin\Phi_q\sin\phi_q)=
\sin\theta_q\cos(\phi_q-\Phi_q), \\   
\hat{y}(\theta_q,\phi_q)&\equiv&\sin\theta_q
(\cos\Phi_q\sin\phi_q-\sin\Phi_q\cos\phi_q)=
\sin\theta_q\sin(\phi_q-\Phi_q), \\   
\hat{z}(\theta_q,\phi_q)&\equiv&\cos\theta_q,
\end{eqnarray}
and $a$, $b$, and $c$ are the axis lengths of the ellipse.  

We have found it best
to fix the axis ratios of the ellipse by computing the maximum extent
of the surface defined by $\tilde{r}(\theta_q,\phi_q)$ in each principal
direction, such that
\begin{eqnarray}
a_0&=&\max |\tilde{r}_{SPH}(\theta_q,\phi_q)\cdot \hat{x}(\theta_q,\phi_q)|,\\
b_0&=&\max |\tilde{r}_{SPH}(\theta_q,\phi_q)\cdot \hat{y}(\theta_q,\phi_q)|,\\
c_0&=&\max |\tilde{r}_{SPH}(\theta_q,\phi_q)\cdot \hat{z}(\theta_q,\phi_q)|.
\end{eqnarray}
Since this prescription can lead to some particles with
$\rho_a>\rho_{min}$ falling outside of the surface, we
multiply the distance in all directions by the smallest factor $F_0$
required to encompass all SPH particles whose density is greater than
$\rho_{min}$, typically leading to an increase in linear size of
no more than $2\%$, such that $a=a_0 F_0$, $b=b_0 F_0$, $c=c_0 F_0$, and
\begin{equation}
F_0=\max\left[\tilde{r}(\theta_q,\phi_q)\sqrt{\frac{\hat{x}^2}{a_0^2}+
\frac{\hat{y}^2}{b_0^2}+\frac{\hat{z}^2}{c_0^2}}\right].
\label{eq:f0}
\end{equation}
A typical surface fit is shown in the upper right panel of
Fig.~\ref{fig:coords}.  The outer two domains are defined in terms of
these new coordinate systems as well, to allow us to match field
values and their derivatives at the boundaries.

Using the collocation points derived from these surfaces, our next
task is to calculate the value of the field equation source terms at
these points.   Since calculating $E$ and $S$ for each SPH particle
from Eqs.~\ref{eq:defn_E} and \ref{eq:defn_S} 
would require a great deal of ultimately needless algebraic work, we do not calculate the
SPH expressions for these quantities at collocation points.
Instead, at every collocation point, we calculate the
SPH expression for the rest mass density, the ``rest pressure'' 
$P_*\equiv k\rho_*^{\Gamma}$, and the density-weighted average
velocity $\tilde{u}_i$.  Using these values, as well as the field
values from the previous iteration, we calculate $\gamma_n$ from Eq.~\ref{eq:gamman}, and then 
$E$ and $S$ at every collocation point, and proceed to solve the field
equations, using the iterative techniques described in GGTMB.  
Typically, we require $\approx 50$ iteration loops to
achieve a solution for which no field value varies by more than 1 part
in $10^9$ from one iteration to the next.

Once we have solved the field equations, we use the spectral expansion
of the fields to calculate as much as we can of the terms in the force
equation, before reading off the values at every particle position,
which takes a significant amount of time.  
In practice, we use the
spectral decomposition to calculate the prefactor for the pressure
force term, $NA^3$, the vector sum of the terms involving derivatives of the
conformal factor and lapse function, the nine first derivatives of the
shift vector, and the radiation reaction terms.  While it would
be faster to calculate the force term involving the derivative of the
shift vector completely in the spectral basis, we have found it to be
inadvisable.  This term alone is linear in the velocity, and it is
inconsistent to use an averaged velocity in this term 
on the RHS of the force equation
to calculate the rate of change of each particle's individual velocity
on the LHS.
Thus, denoting terms calculated within the spectral basis and exported
to particle positions
by ``sb'', and those calculated 
using SPH techniques only by ``SPH'', Eq.~\ref{eq:dutildedt} is
truly evaluated as
\begin{equation}
\frac{d\tilde{u}_i}{dt}=[-NA^3]_{sb}\left[\frac{\nabla_i
P}{\rho_*}\right]_{SPH}
+\left[\frac{Nh({\gamma_n}^2-1)}{A\gamma_n}\nabla_i A -h\gamma_n\nabla_i N\right]_{sb}
-[\nabla_i N^j]_{sb}[\tilde{u}_j]_{SPH}.
\end{equation}
After calculating the forces on each particle and advancing the
velocities by a full timestep, we are still left with the task of
recomputing Eq.~\ref{eq:uv}, which relates $v^i$ and $\tilde{u}_i$.
Since the sources for the field equations are velocity dependent, and
we have just advanced $\tilde{u}$, we rerun the field solver with
the new values of $\tilde{u}_i$, and compute Eq.~\ref{eq:uv} in the form
\begin{equation}
v^i=[N^i]_{sb}+\left[\frac{1}{A^2hu^0}\right]_{sb}[\tilde{u}_i]_{SPH}.
\end{equation}
Finally, we record the GW strains, ADM mass and system
angular momentum,
after rotating all positions, velocities, and accelerations back
to the inertial frame by means of the inverse Euclidean transformation
to the one at the beginning of the iteration

There are several different ways to calculate the ADM mass numerically,
all of which should be equivalent to Eq.~\ref{eq:admmass}, yielding 
an important check on the code.  
First, we calculate the system's ADM mass using by
taking a surface integral at spatial infinity (see Eq.~65 of GGTMB),
\begin{equation}
M_0=-\frac{1}{2\pi} \oint_\infty \partial^i A^{1/2} dS_i.
\label{eq:adm_sb}
\end{equation}
This quantity can be compared to the particle-based expression, with two
important caveats.  Since the extrinsic curvature contribution to the
ADM mass, Eq.~\ref{eq:rho2},
does not have compact support, there is no way to convert the
integral into a sum over particles that have a
different spatial extent.  Second, noting our concerns about
exporting large numbers of terms from the spectral representation to
particle positions, we perform some of the algebra involved in
determining the ADM mass in the spectral representation.  In the end,
the particle-based expression for the ADM mass becomes
\begin{equation}
M_0=(\int \rho_2 d^3x)_{sb}+
\sum_a m_a\left[\frac{\rho_1}{\rho_*}\right]_{sb},
\label{eq:adm_sph}
\end{equation}
and we define $\tilde{\rho}\equiv\rho_1/\rho_*$ as the ratio of ADM mass
density to rest mass density, calculated using LORENE techniques at
all SPH particle positions.

In a similar fashion, there are two ways to calculate the system angular
momentum.  We check the behavior of $J$ in the spectral basis (see
Eq.~67 of GGTMB), 
\begin{equation}
J_i=\frac{1}{16\pi}\epsilon_{ijk}\oint_\infty (x^jK^{kl}-x^k K^{jl}) dS_l,
\label{eq:j_sb}
\end{equation}
as well as the SPH expression for the angular momentum,
\begin{equation}
J_i=\epsilon_{ijk}\sum_a m_a x^j \tilde{u}_k.
\label{eq:j_sph}
\end{equation}

Since the CF formalism is time-symmetric, the dissipative effects of
gravitational radiation backreaction have to be added in by hand, just
as they are for PN calculations. 
Previous PN calculations of binary NS systems (FR and \cite{APODR}) have
typically employed the exact 2.5PN formulae introduced by \cite{BDS} 
to describe lowest-order GW losses from the system.  Unfortunately,
those equations are not applicable to CF calculations, since they are
written in terms of fields defined in the PN approximation that
differ from those defined in the CF approximation.  For this work, we
follow the approach of \cite{WMM}, using the slow-motion
approximation to estimate the radiation reaction potential of the
system.  While this method contains some obvious flaws, most
obviously the fixed spatial dependence of the
radiation reaction potential, it does yield a
backreaction force which is quantitatively correct in overall magnitude. 
These approximations should not affect our calculations to a large
degree.  While the infall velocity of the binary prior to plunge is
driven by the GW backreaction, a different
regime occurs after dynamical instability sets in.
During this period, the evolution is almost completely hydrodynamic in
nature \cite{RS1}.  While the chosen GW backreaction treatment 
may affect the final
mass and angular momentum of the resulting merger remnant, it will
play only a secondary role in the detailed evolution of the fluid
configuration, since GW backreaction 
become less important during the coalescence. 

From Eq.~51 of \cite{WMM}, the radiation reaction force in the
slow-motion approximation is given by adding a term,
\begin{equation}\label{eq:rrforce}
a_{i:reac}=N^2hu^0\nabla_i \chi,
\end{equation}
to the RHS of Eq.~\ref{eq:dutildedt}.  
We define the radiation reaction potential
$\chi$ in a similar but slightly different way than their Eq.~56,
such that
\begin{equation}
\chi=\frac{1}{5}x^k x^l Q^{[5]}_{kl},
\end{equation}
similar to the approach taken in \cite{Oech}.  As they do, we define
the quadrupole moment as
\begin{equation}
Q_{kl}=STF\left[\int \rho_{ADM} x_k x_l d^3 x\right],
\end{equation}
noting that our ``$\rho_{ADM}$'' corresponds to many other authors'
$S_{\psi}$, or some multiple thereof.  The
expression ``STF'' refers to the symmetric, trace-free component of the
tensor, which is linked to the gravitational radiation production in
the quadrupole limit.
As noted previously in our calculations of the system's ADM mass, we can 
evaluate the contribution from $\rho_1$ using standard SPH summation
techniques, but 
the second term can only be found using LORENE integration techniques.
Thus,
\begin{equation}
Q_{kl}\equiv Q_1+Q_2={\rm STF}\left[\left(\sum_a m_a \bar{\rho}_a
x^k_a x^l_a\right) +
\left(\int  d^3x \rho_2 x^k x^l\right)_{sb} \right],
\end{equation}
where $Q_1$ and $Q_2$ reflect the contributions from 
$\rho_1$ and $\rho_2$, respectively.
Using SPH, we can take the first time derivative of the former half, so
long as we ignore the Lagrangian 
derivative $d\rho_1/dt$, which should be
essentially negligible during our calculations.  We find
\begin{equation}\label{eq:q1}
(\dot{Q}_1)_{kl}=STF\left[\int \rho_1 (x_k v_l+x_l v_k) d^3 x \right].
\end{equation}
To calculate the rate of change of the extrinsic curvature, we assume
that the time
variation in the tensor is due solely to the orbital motion
(rather than an overall change in magnitude of the tensor components in
a corotating frame), which yields 
\begin{eqnarray}
(\dot{Q}_2)_{xx}=-(\dot{Q}_2)_{yy}&\approx&2\omega({Q}_2)_{xy}, \\
(\dot{Q}_2)_{xy}&\approx&2\omega\left[\frac{({Q}_2)_{xx}-({Q}_2)_{yy}}{2}\right],\\
\end{eqnarray}
where the factor $2\omega$ reflects the fact that the quadrupole
tensor makes two cycles during every orbital period.
To calculate the fifth time derivative of the quadrupole tensor, we
use the same technique with both components of the tensor,
finding
\begin{equation}
Q^{[5]}_{kl}\approx 16\omega^4 \dot{Q}_{kl},
\end{equation}
where in all cases the system's instantaneous angular velocity is calculated 
as the ratio of the angular momentum to the moment of inertia, 
\begin{equation}
\omega\equiv\frac{\sum_a m_a (x_a (v_y)_a - y_a (v_x)_a)}{\sum_a m_a (x_a^2+y_a^2)},
\label{eq:omega_sph}
\end{equation}
which holds exactly in the quadrupole limit for synchronized binaries.

Calculating the GW signal and energy spectrum is a more straightforward
task which can be done after the calculation is finished.  We
calculate the GW strain in two independent polarizations for an
observer located at a distance $d$ from the system perpendicular to
the orbital plane from the lowest-order quadrupole expressions,
\begin{eqnarray}
dh_+&=&Q^{[2]}_{xx}-Q^{[2]}_{yy}, \label{eq:hplus}\\
dh_\times&=&2Q^{[2]}_{xy},\label{eq:htimes}
\end{eqnarray}
where we calculate the second time derivatives of the quadrupole
tensor by numerically differentiating the results from
Eq.~\ref{eq:q1}.  In terms of the Fourier transform of the quadrupole
moment,
\begin{equation}
\tilde{Q}^{[2]}_{kl}(f_{GW})\equiv\int e^{2\pi ift}Q^{[2]}_{kl}(t)dt,
\end{equation}
where $f_{GW}\equiv 2f_{orb}$ is the {\it GW}
frequency, the GW energy spectrum is computed as \cite{ZCM1}
\begin{eqnarray}
\frac{dE}{df_{GW}}&=& \pi f_{GW}^2 \left(\frac{8}{15}\left[ 
(\tilde{Q}^{[2]}_{xx}-\tilde{Q}^{[2]}_{yy})^2+
(\tilde{Q}^{[2]}_{xx}-\tilde{Q}^{[2]}_{zz})^2+
(\tilde{Q}^{[2]}_{yy}-\tilde{Q}^{[2]}_{zz})^2\right]+\right.\nonumber\\
& &\;\;\;\;\;\;\;\;\;
\left.\frac{8}{3}\left[(\tilde{Q}^{[2]}_{xy})^2+(\tilde{Q}^{[2]}_{xz})^2+
(\tilde{Q}^{[2]}_{yz})^2\right]\right).\label{eq:dedf}
\end{eqnarray}

\subsection{Merging systems}\label{sec:merging}

As the stars in the binary system spiral inward, they reach a point where the
density distributions begin to overlap as matter from the inner edge
of each star falls onto the surface of the other.  Our field solver
can handle this situation, since it does not assume that the matter
sources are spatially distinct, but the surfaces required to envelop
the particles from each star would become poorer and poorer fits to
the two NS.  Noting this, we alter
the approach described above in a number of ways when the particles
first cross through the inner Lagrangian point at the center of the
system, in such a way that the surfaces we define for each star always
remain smooth and still do an acceptable job of describing the true
density distribution in a meaningful way.  

Once the binary separation shrinks sufficiently, 
matter streams from the inner edge of each
NS toward the other NS, flowing along the
surface of the companion.  These counter-streaming, low-density flows
lead to the formation of a vortex sheet 
(FR3).   Mass transfer typically occurs 
before the triaxial ellipsoidal surfaces
used to define the two stars overlap, since particles crossing from
one star to another generally fall beneath the density cut used to
define each surface.  
To account properly for the star to which each particle is bound,
we declare particle ``a''  a member of the first star if $x'_a<0$ and
a member of the second star if $x'_a>0$, at least for the purposes of
defining each star's center of mass and tidal lag angle, 
as described in Eqs.~\ref{eq:phiprime}--\ref{eq:f0}.

This approach entails further alterations 
once the ellipsoidal computational domains from each star
begin to overlap.  Using the $y'$-axis as the dividing line between
the two stars is fine for determining the center-of-mass, tidal lag
angle, and ellipsoidal surface for each star, but it is inappropriate
to draw a fixed line at $x'=0$ when calculating field
equation source terms.  These would induce 
a sudden density drop from finite density
at small negative values of $x'$ to zero density at positive $x'$
values (for the first star, vice versa for the second), 
leading to large Gibbs effects.  It is equally
inappropriate to count one particle as a member of both stars,
since we would end up double-counting its density contribution to the
field equations.  Instead, we introduce a weight function $f_a\equiv 
f(\vec{x}'_a)$, 
for each particle in the overlapping region in
such a way that $f=0$ at the surface of star 1, $f=1$ on the surface
of star 2, and $0<f<1$ in a 
spatially differentiable way within the overlap region.  
To do so, we define $f_1$ and $f_2$, the fractional squared distance 
outward a point lies from the center of each star to the surface.  For
the first star,
\begin{equation}
f_q(x')\equiv \frac{(X'_q)^2}{a^2}+\frac{(Y'_q)^2}{b^2}+
\frac{(z')^2}{c^2},\label{eq:fq}
\end{equation}
where 
\begin{eqnarray}
X'_q\equiv x_q\cos\Phi_q+y'\sin\Phi_q,\label{eq:bigxprime}\\
Y'_q\equiv-x_q\sin\Phi_q+y'\cos\Phi_q,\label{eq:bigyprime}
\end{eqnarray}
are the rotated coordinates used to
define the (tidally lagging) ellipsoidal surface of the star, and
$x_q(x')\equiv x'-x'_q$ is the distance in the $x'$-direction from the
center-of-mass of star $q$.  A picture of these quantities is shown in
the bottom left panel of Fig.~\ref{fig:coords}.
In terms of $f_q$, we define our 
overlap function $f_a$ for each particle such that
\begin{equation}
f_a=f(\vec{x}'_a)=\frac{(1-f_1)^3}{(1-f_1)^3+(1-f_2)^3}.
\end{equation}
For the first star, source terms in the field equations are evaluated as
\begin{eqnarray}
(\rho_*)_1&=&\sum_a m_a f_a W_{a},\\
(P_*)_1&=&\frac{(\rho_*)_1}{(\rho_*)_1+(\rho_*)_2}k[(\rho_*)_1+(\rho_*)_2]^{\Gamma},\\
(\tilde{u}_i)_1&=&\frac{\sum_a m_a f_a (\tilde{u}_i)_a W_{a}}{\sum_a m_a f_a
W_{a}},
\end{eqnarray}
where $W_a$ is the smoothing kernel function evaluated between the
collocation point and overlapping particles.  Complementary expressions
hold for the second star by substituting $(1-f_a)$ for $f_a$.

This approach allows us to calculate the fields properly using
spectral methods well into the merger, but a final modification is
necessary to bring us to the point where the density profile of the
matter can better be described by viewing it as a single object.
When the inner edge of one star overlaps the center of the other
star, the density profile typically becomes bimodal, leading
to spurious results from the spectral expansion.  
To guard against this happening, we use a
relatively simple approach.  If the surface of one ellipse extends more
than halfway from $x'=0$ toward the center-of-mass of the other star
at $x'_q=\pm R/2$, we linearly rescale all surface points that lie
across the $y'$-axis.  Thus, defining
$x_q(\theta_q,\phi_q)=r(\theta_q,\phi_q)\sin\theta_q\cos\phi_q$ in
accordance with our previous notation,
if the
surface of either star extends to a maximum value
$x_{q:max}\equiv\max(|x_q|)> 3R/4$, (or
in other words, if
$\max(|r(\theta_q,\phi_q)\sin\theta_q\cos\phi_q-x'_q|)> R/4$,)
we adjust all points on the other side of the y'-axis, yielding
\begin{equation}
x_{q:new}(\theta_q,\phi_q)=\frac{R}{2}+\frac{R}{4}\left(
\frac{x_{q:old}(\theta_q,\phi_q)-R/2}{x_{q:max}(\theta_q,\phi_q)-R/2}\right)
~~{\rm for~all~points~with~}x_q(\theta_q,\phi_q)> R/2.
\end{equation}
To evaluate the weight function, we adapt the x'-dependence
correspondingly, such that for particles with $x_q>R/2$
\begin{equation}
f_{a:new}(\vec{x}_q)=f_{a:old}(\kappa\cdot\vec{x}_q),
\end{equation}
where the rescaling factor $\kappa\equiv
R/2+(x_q-R/2)\frac{x_{q:max}-R/2}{R/4}$. 
The bottom right panel of Fig.~\ref{fig:coords} demonstrates this last
coordinate transformation.

Eventually, the system will reach a point where it can no longer be
properly described as a binary, and our field solver fails to
converge.  Before this happens, we reach a point where we can describe
the system as a rapidly rotating single star,
using all the methods described above for computing the evolution, but
now assuming that all particles comprise the same star.  We have found
that our results are independent of the exact moment at which we make
this conversion.  In the following discussion, we will show the
results for runs where we perform the conversion at the earliest
possible time for which the field solver will converge to a solution
when the matter configuration is treated as a single star.

\section{Test Calculations}\label{sec:tests}

We have performed several tests to check the accuracy and numerical
stability of our code, for both single-star and binary systems.  We
have studied the behavior of the code for 
spherically symmetric problems for which the exact field
solution can be calculated semi-analytically: the Oppenheimer-Volkov
solution for a static spherical star and the collapse of a
pressureless dust cloud initially at rest.  We have also calculated
the dynamical evolution of the binary quasi-equilibrium models
calculated 
in TG, without the inclusion of radiation reaction
forces.  A summary of all our calculations, including those used for
testing the code, can be found in Table~\ref{table:runs}.

\subsection{The Oppenheimer-Volkov (OV) Test}\label{sec:ov}

Since the CF formalism is exact for spherically symmetric systems,
which can always be described in isotropic coordinates, it
is fair to expect that any working code should be able to reproduce
the well-known Oppenheimer-Volkov solution exactly, noting that the
traditional form of the OV solution needs to be rewritten into CF
coordinates.  With $\rho'\equiv\rho(1+\epsilon)$ as the total energy
density (including both the rest and internal energy densities), the
OV equations are typically written
\begin{eqnarray}
\frac{dm}{d\bar{r}}&=&4\pi \bar{r}^2\rho',\\
\frac{dP}{d\bar{r}}&=&-\frac{(\rho'+P)
\left(m+4\pi P\bar{r}^3\right)}{\bar{r}^2-2m\bar{r}},\\
\frac{d\Phi}{d\bar{r}}&=&\frac{\left(m+4\pi
P\bar{r}^3\right)}{\bar{r}^2-2m\bar{r}},
\end{eqnarray}
for an interior metric in the form
\begin{equation}
ds^2=-e^{2\Phi}dt^2+\left(1-\frac{2m}{\bar{r}}\right)^{-1}d\bar{r}^2
+\bar{r}^2d\Omega^2,
\end{equation}
and exterior metric in the Schwarzchild form
\begin{equation}
ds^2=\left(1-\frac{2M_0}{\bar{r}}\right)dt^2+
\left(1-\frac{2M_0}{\bar{r}}\right)^{-1}d\bar{r}^2+\bar{r}^2d\Omega^2,
\label{eq:sch_ex}
\end{equation}
where the star's total gravitational (ADM) mass 
$M_0\equiv m(\bar{r}_s)$, and $\bar{r}_s$ is
the Schwarzchild coordinate radius of the stellar surface.
A quick comparison with the CF metric, Eq.~\ref{eq:metric},  
shows that $N=e^{\Phi}$ inside the star, but the
conversion between $A$ and $\Phi$ requires more care, since we have to
determine the relationship between the two coordinate radii
$r(\bar{r})$.  The simplest way to determine the change of coordinates
is to solve for the values of $r_s$ and $\bar{r}_s$, the radii of the
stellar surface, using the asymptotic behavior of the exterior
solution (following the same logic used in exercise 31.7 of
\cite{MTW}).  
Comparing the radial and angular parts of the metric, we find
\begin{eqnarray}
A^2 dr^2 &=& \left(1-\frac{2M_0}{\bar{r}}\right)^{-1} dr^2, \\
A^2 r^2 &=& \bar{r}^2. \label{eq:rrbar}
\end{eqnarray}
Dividing and taking a square root, we find
\begin{equation}
\frac{dr}{r} = \frac{d\bar{r}}{\sqrt{\bar{r}(\bar{r}-2M_0)}}.
\end{equation}
Integrating yields
\begin{equation}
\ln r+k = 2\ln \left(\sqrt{\bar{r}}+\sqrt{\bar{r}-2M_0}\right),
\end{equation}
and we find
\begin{equation}
kr=\left(\sqrt{\bar{r}}+\sqrt{\bar{r}-2M_0}\right)^2=
2\bar{r}-2M_0+2\sqrt{\bar{r}(\bar{r}-2M_0)}.  
\end{equation}
The asymptotic behavior at infinity indicates that we must have
$k=4$, so our final expression for $r(\bar{r})$ takes the form
\begin{equation}
r=\frac{1}{2}\left(\bar{r}-M_0+\sqrt{\bar{r}(\bar{r}-2M_0)}\right).
\end{equation}
For a star with Schwarzchild surface radius $\bar{r}_s$, we find that the CF
surface radius is given by
\begin{equation}
r_s=\frac{1}{2}\left(\bar{r}_s-M_0+\sqrt{\bar{r}_s(\bar{r}_s-2M_0)}\right),
\end{equation}
the surface value for the conformal factor is
\begin{equation}
A_s=\frac{\bar{r}}{r}=\frac{2\bar{r}_s}{\bar{r}_s-M_0+\sqrt{\bar{r}(\bar{r}-2M_0)}},
\end{equation}
and the lapse function at the surface is given by
\begin{equation}
N_s=\sqrt{1-\frac{2M_0}{\bar{r}_s}}.
\end{equation}
To solve for the interior metric, we add an equation to the OV set to
take into account the different radial schemes, defining a scale free
conformal radius satisfying a boundary condition 
$\lim_{\bar{r}\rightarrow 0} r_0=\bar{r}$ whose radial behavior is
given by
\begin{equation}
\frac{dr_0}{d\bar{r}}=\frac{r_0}{\sqrt{\bar{r}(\bar{r}-2m)}}.
\end{equation}
This yields an expression for $r_0(\bar{r})$ which is defined up to an
arbitrary multiplicative constant $k$, such that $r_0=k\cdot
f(\bar{r})$, where $f$ is determined from the mass distribution.
To find $r(\bar{r})$, we merely
set $k=r_s/(r_0)_s$, which implies $r=\bar{r}r_s/(r_0)_s$ 
and determine $A$ from Eq.~\ref{eq:rrbar}.

To test the code, we ran three calculations whose results could be
compared with well-known semi-analytic solutions.  First,
we constructed an equilibrium model for an
isolated NS, whose density profile was given by an OV solution for a
$\Gamma=2$ EOS, with unit ADM mass, and a conformal radius $r_s=6.874$.
The solution has a total baryonic mass $M_b=1.066$, and an areal
radius $\bar{r}_s=7.913$.
We started run OV1 by taking this solution as an initial condition, 
and used it to test the overall
stability of the code for equilibrium configurations.   Next, we constructed
a similar model with the same mass and EOS,
but scaled to an initial radius $10\%$ larger.  Run OV2 is a
dynamical calculation started from this initial condition using our
standard evolution code, enabling us to study the oscillations
around our equilibrium solution.  Last, we took the non-equilibrium
configuration from run OV2, but added a ``relaxation'' drag term of the
form $-\tilde{u}_i/t_{relax}$ to the RHS of
the force equation, Eq.~\ref{eq:dutildedt}, with $t_{relax}/M_0=7.9$.
This provides an overdamped force for run OV3 since the dynamical timescale
$t_D/M_0\equiv (G\rho)^{-0.5}=18$.
All three runs were followed until $t/M_0=120$, corresponding to
$6.7$ dynamical times for the NS, and in all cases radiation reaction
was turned off.

\begin{figure}
\centering\includegraphics[width=6in]{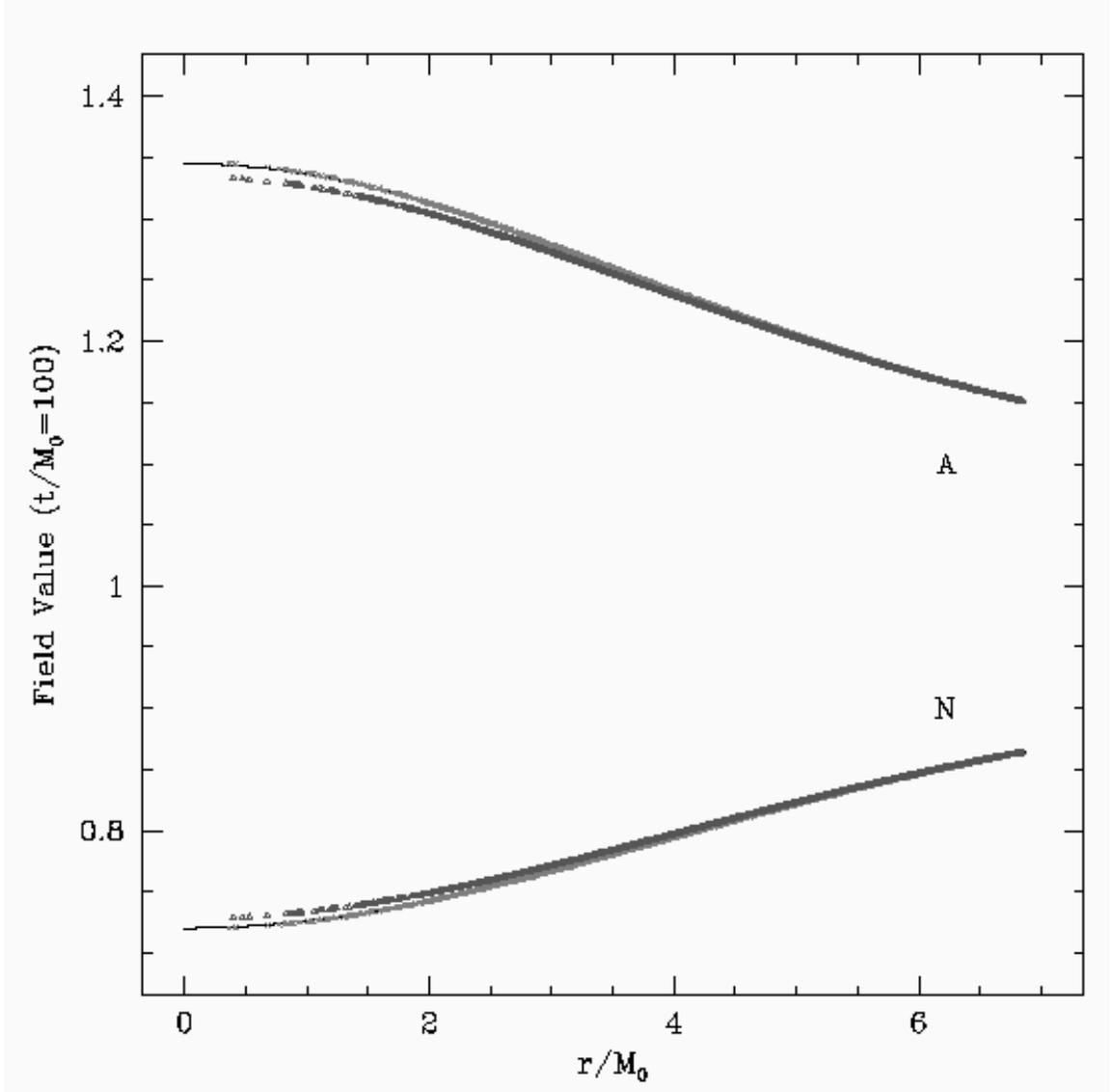}
\caption{Conformal factor $A$ and lapse function $N$ for an
  Oppenheimer-Volkov solution with a $\Gamma=2$ polytropic EOS, and
  conformal radius $r_s/M_0=6.874$  (solid lines),
  compared to our computed values at $t/M_0=100$ 
for two models: run OV1 started from
  equilibrium (crosses) and run OV3 started from a configuration 
10\% larger with
  a velocity damping term used to drive the system toward equilibrium
  (triangles).  The agreement is within $1\%$ for all particles.
   Units are defined such that $G=c=1$.  All masses, lengths, and
  times in the OV and dust cloud calculations are made dimensionless
  by scaling results against the system's initial gravitational (ADM)
  mass.  Note that the conformal radius is not equivalent to the
  areal radius typically used in solving the OV equation.}
\label{fig:ov_an}
\end{figure}

\begin{figure}
\centering\includegraphics[width=6in]{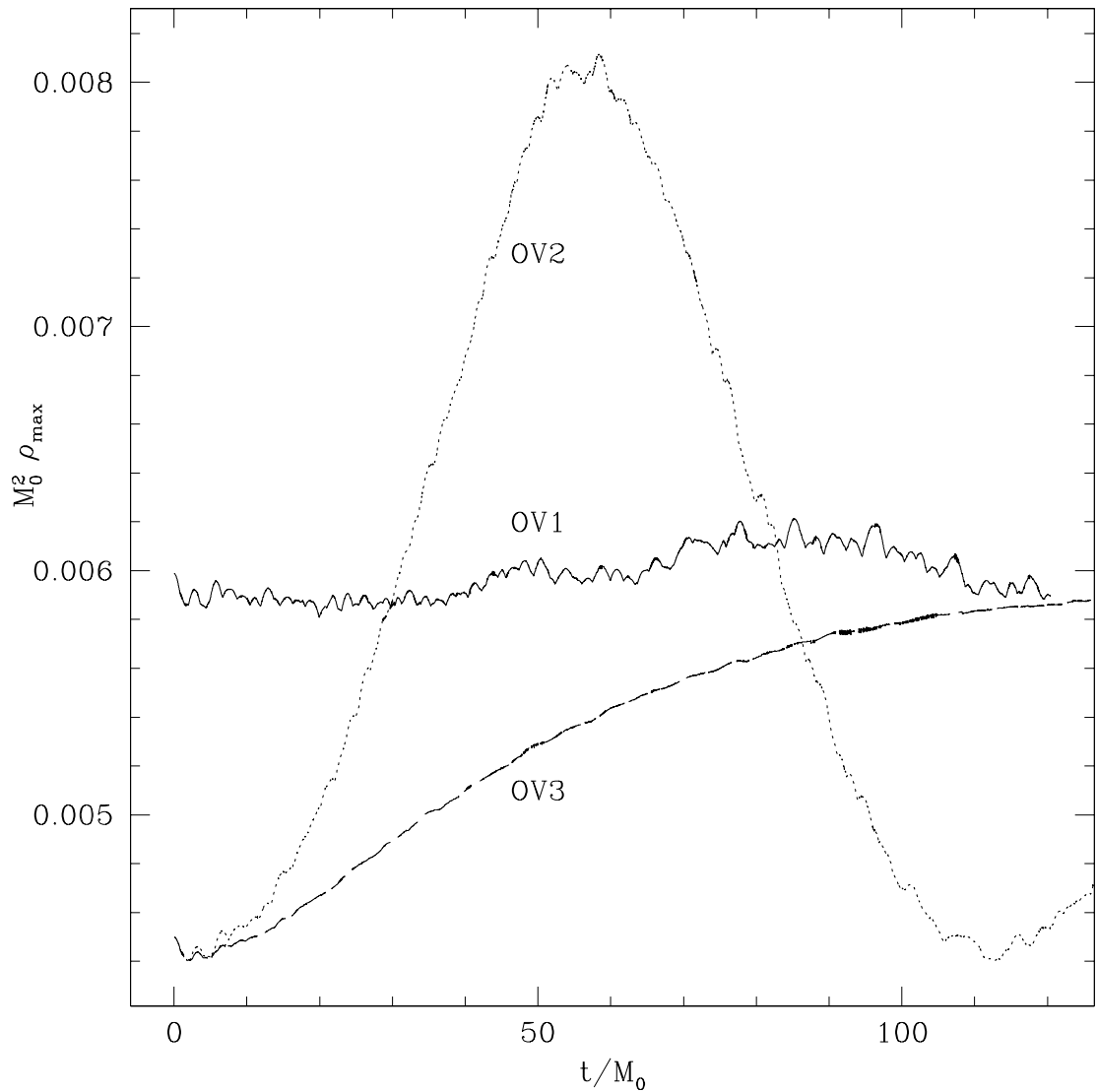}
\caption{Evolution of the maximum density for three runs based on the
  OV solution described in Fig.~\protect\ref{fig:ov_an}.  Run OV1 (solid
  line) was started from equilibrium, and shows only small variations
  which result from typical uncertainties in SPH summation.  Run OV2
  (dotted line)
  used the same EOS, but was started from a radius $10\%$ smaller than
  the equilibrium value, showing sinusoidal oscillations with a
  period of $T/M_0=112$ and no systematic drift.  Run OV3
  (dashed line) was started from the same configuration as run B, but
  with an overdamped drag term to force the system toward equilibrium,
  converging rapidly toward the proper value of $M_0^2 \rho_{max}=0.006$.}
\label{fig:ov_rhomax}
\end{figure}

In Fig.~\ref{fig:ov_an}, we show the radial profiles of the lapse function
and conformal factor at $t/M_0=100$ for runs A and C, along with the
correct OV solution.  It is no surprise that
run A has essentially remained the same, since the
initial field values were essentially exact, but it is reassuring that
run C has converged as well toward the same solution.  Indeed, results for
configurations at later times continue to converge toward the exact solution.
In Fig.~\ref{fig:ov_rhomax}, we show the evolution of the maximum
central density for the three runs, as well as the predicted value
from the OV solution.  The maximum density from run A stays near this
value throughout the evolution, with small deviations which result
from the unavoidable discretization effects present in SPH; in
general, SPH will yield very accurate global integrals over a mass
distribution, since numerical noise smooths out, but demonstrates
significant noise in quantities defined for individual particles,
which vary iteration to iteration as each particle's neighbor list
adapts to current conditions.  The maximum density for run B
oscillates around the proper value with a period of $T/M_0=112$,
showing no signs of systematic drift. This is very close to the proper
value for the limiting case of infinitesimal radial variations,  
$T/M_0=104$, which we find by interpolating from the values given in
Table~A18 of \cite{KR}, after scaling their results to our units. 

\subsection{Spherical Dust Cloud Collapse}\label{sec:dust}

To further test the dynamical aspects of the code, we computed the
evolution of a uniform density dust cloud, i.e., a spherical
distribution of matter with zero pressure, started initially from
rest.  This is a familiar problem from cosmology, and the solution is
well-known, but the conversion to CF coordinates and the
matching conditions at the surface of the dust cloud make the
resulting expressions considerably more complicated.  The complete
description of the metric as a function of time was derived 
for the case of both maximal time-slicing
\cite{Petrich}
and polar time-slicing \cite{Petrich2}; it is the former case which can
be compared to our results.  In this approach, the field values and CF
time and position variable values are
computed by solving a set of ordinary differential equations which
describe the behavior of the fields in terms of the comoving time and
space variables.  Using these, it is a simple process of interpolation
to derive the fields as functions of CF time and position.  We checked
our integration code by comparing our results against the plots in
\cite{Petrich}, finding perfect agreement.

To construct our initial configuration, we used essentially
the same techniques described above.
Particles were laid out in an
HCP lattice, with masses set proportional to the analytically known
value of $\rho_*(r)$, which we derive as follows.  Note that $\rho_*$
varies with radius; it is $\rho(r)$ that is initially
uniform.

The total mass of the cloud was set to unity, and the initial radius
in comoving coordinates to $\bar{r}_s$, the parameter used for all
figures in \cite{Petrich}.  The initial metric in
comoving coordinates, for a cloud with unit mass and comoving radius
$\bar{r}_s$, is given by
\begin{eqnarray}
ds^2&=&-d\tau^2+a(\tau)^2(d\chi^2+\sin^2\chi d\Omega^2),\\
a(\tau=0)&\equiv& 2/(\sin \chi_s)^3, \\
\bar{r}_s&\equiv&2/(\sin \chi_s)^2,
\end{eqnarray}
with the exterior metric described by the Schwarzchild form,
Eq.~\ref{eq:sch_ex}, with $\bar{r}=a(\tau=0)\sin\chi$.  
In terms of $\chi_s$, the CF radius is given by
\begin{eqnarray}
r_s=\frac{1}{2}\left(\bar{r}_s-M_0+\sqrt{r_s(r_s-2M_0)}\right)&=&\frac{1}{2}
\left(\frac{2}{\sin^2\chi_s}-1+\sqrt{\frac{4}{\sin^4\chi_s}-
\frac{4}{\sin^2\chi_s}}\right)\nonumber\\
&=&\frac{1}{2}\frac{1+\cos\chi_s}{1-\cos\chi_s}. \label{eq:rchi}
\end{eqnarray}
Equating interior metric coefficients, we set $Adr=ad\chi$ and
$Ar=a\sin\chi$, dividing and then integrating to find
\begin{eqnarray}
r&\propto&\left(\frac{1-\cos\chi}{1+\cos\chi}\right)^{1/2},\\
&=&\frac{1}{2}\left(\frac{1-\cos\chi}{1+\cos\chi}\right)^{1/2}
\left(\frac{1+\cos\chi_s}{1-\cos\chi_s}\right)^{3/2},
\end{eqnarray}
where the proportionality constant is determined from Eq.~\ref{eq:rchi}. 
The initial conformal factor can now be written down, since
\begin{equation}
A=\frac{a\sin\chi}{r}=\frac{4(1+\cos\chi)}{(1+\cos\chi_s)^3},
\end{equation}
and we can simplify the resulting expression by noting that
$\cos\chi_s=(2r_s-1)/(2r_s+1)$ and
$\cos\chi=(1-r^2/2r_s^3)/(1+r^2/2r_s^3)$.
Since the matter starts from rest, we know that $u_i=0$ initially, and
thus $\gamma_n=1$ everywhere, and we find the initial rest-mass
density profile is given as a function of radius by
\begin{equation}
\rho_*=\gamma_nA^3(\frac{3M}{4\pi \bar{r}_s^3})=\frac{3}{4\pi r^3}\left(\frac{1+\frac{1}{2r_s}}{1+\frac{r^2}{2r_s^3}}\right)^3.
\end{equation}

Once the initial particle configuration was set, we calculated the
dynamical evolution of the system using the same techniques described
above for the OV case.  Since the pressureless material had no
outwardly directed force, the inevitable fate of the system was
collapse to a BH.

\begin{figure}
\centering\includegraphics[width=6in]{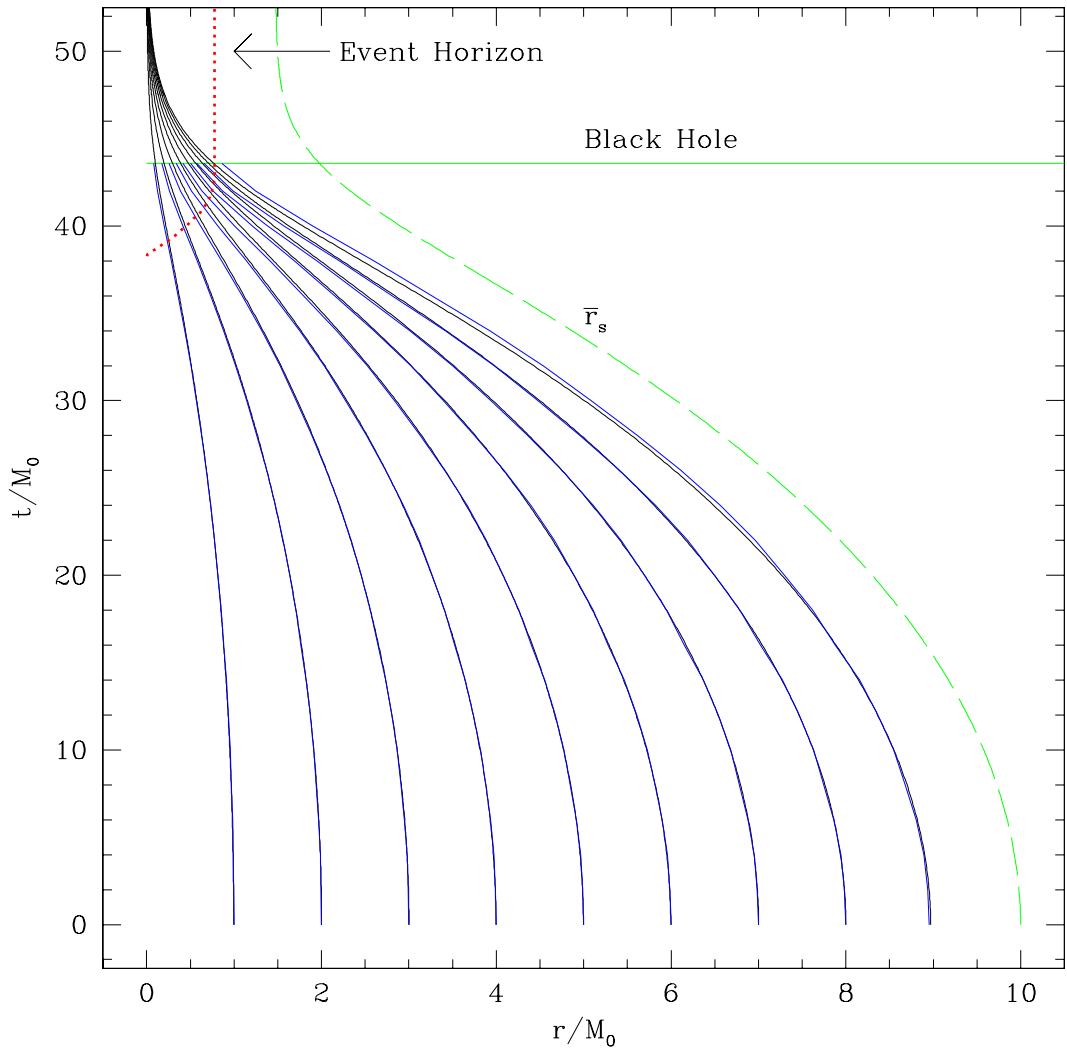}
\caption{A comparison between the actual paths traced out by a set of
equally-spaced Lagrangian tracers in run DC10, our calculation of a collapsing
dust cloud with unit ADM mass and initial areal radius $\bar{r}_s=10$, 
(dashed lines) 
and the exact semi-analytical solution of \protect\cite{Petrich},
shown as solid lines.  All radii are shown here in conformal coordinates.
We see excellent agreement up until the point
where an event horizon forms at the center of the cloud, at
$t/M_0=38$.  The event horizon moves outward (heavy dotted line), 
eventually enclosing the
entire cloud at $t/M_0=43.4$, shown as a horizontal line in the figure.
At this point, when the matter can be properly defined as a BH, 
our field solver stops converging.  For comparison with Fig.~9
of \protect\cite{Petrich}, we also show, as a long-dashed line, the
exact solution for the cloud's surface in comoving (areal) radii, 
$\bar{r}_s(t)$.}
\label{fig:petr_lag}
\end{figure}

To allow for direct comparison with the figures shown in \cite{Petrich}, we
computed the evolution of a dust cloud with unit ADM mass and 
an initial areal radius
$\bar{r}_s=10$, just as they did.
In Fig.~\ref{fig:petr_lag}, we show 
the evolution of a set of equally spaced Lagrangian
tracer particles compared to the exact semi-analytic solution we
computed.  This corresponds to their Figs.~8-9, with two
slight differences.  Our figures are plotted in CF coordinates, rather
than comoving coordinates, and our tracers are equally-spaced in
radius, not in increments of enclosed mass.  For comparison with
their Fig.~9, we also show 
the more familiar areal radius of the cloud, which roughly satisfies
the relation $r_s\approx \bar{r}_s-1$ initially. 
We see the agreement between our calculation and the exact solution is
very good throughout the evolution, up until $t/M_0\approx 38$, where a
slight discrepancy begins to develop, primarily at the surface of the
cloud.  This time corresponds closely
with the formation of an event horizon for the system, shown as a
dotted line, starting at the center at $t/M_0=38$ and moving outward,
reaching the surface of the cloud at $t/M_0=43.4$, shown as a horizontal
line.  This late time discrepancy has two sources.
The first is that SPH, which by definition produces a differentiable
density field, cannot reproduce the step-function density drop at the
surface of the dust cloud.  This explains to a large degree why the
outermost tracers diverge furthest from their exact path.  In
addition, the large field values and gradients found around the event
horizon present a challenge for our field solver.  We typically see a
dramatic increase during this period in the number of relaxation
iterations required to converge to a sufficiently accurate solution.

\begin{figure}
\centering\includegraphics[width=6in]{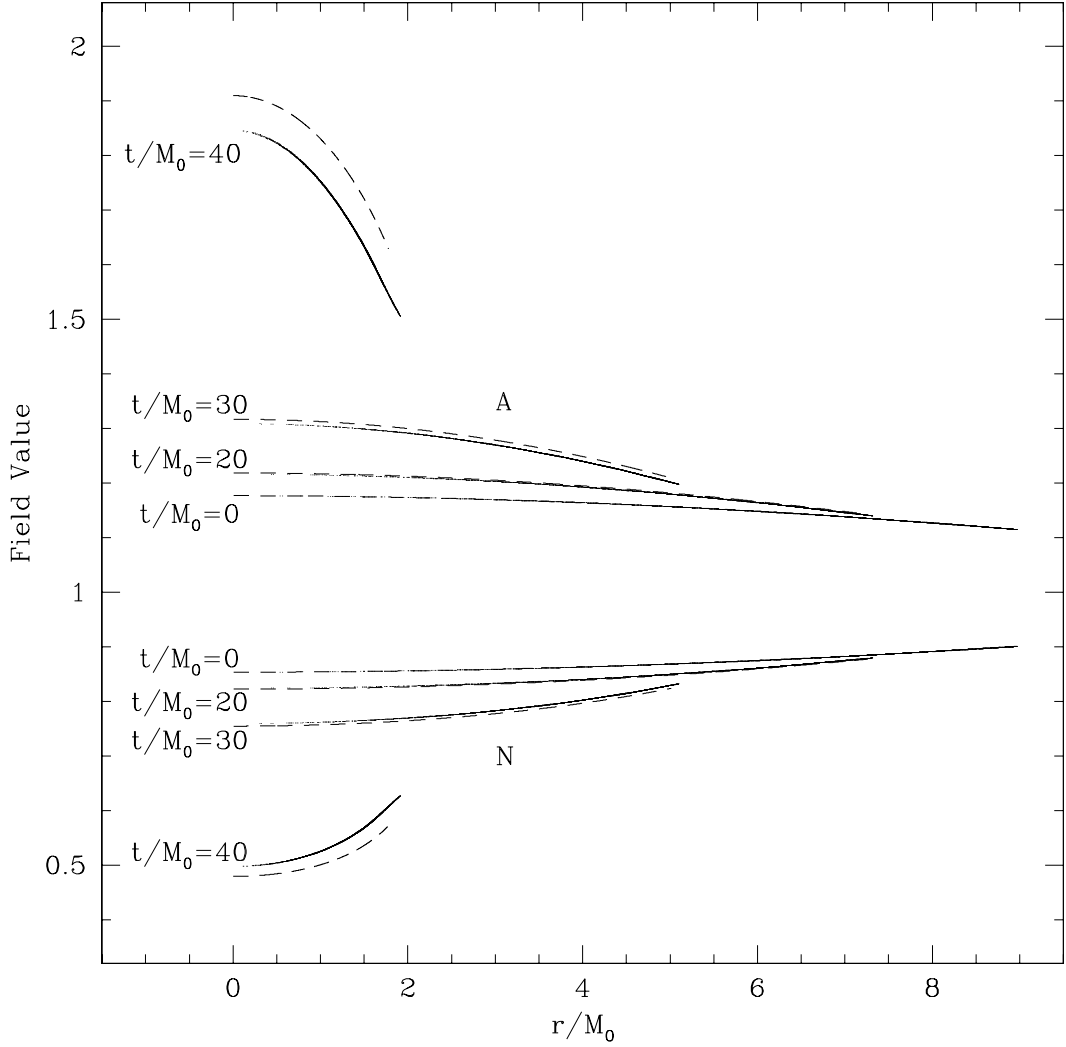}
\caption{The evolution of the conformal factor $A$ and Lapse function
$N$ for the dust cloud in run DC10, 
compared to the exact solution.  We see, at $t/M_0=0$,
$20$, $30$, and $40$, the SPH particle values for the lapse (the four
curves with values less than 1.0) and conformal factor (values greater
than 1.0), shown as points, and the exact solutions, shown as dashed
lines.  The agreement is good up until $t/M_0=30$, but at $t/M_0=40$ we see
some quantitative disagreement, since the field solver breaks down as
we near the point where the cloud collapses completely into a BH.} 
\label{fig:petr_an}
\end{figure}

In Fig.~\ref{fig:petr_an}, we show the evolution of the conformal
factor $A$ and lapse function $N$ for the dust cloud, again in
comparison to the exact solution, at times $t/M_0=0$, $20$, $30$, and
$40$.  We see again that our code can reproduce the proper solution
past $t/M_0=30$, but by $t/M_0=40$ shows non-trivial deviations from the
proper solution.  At $t/M_0=40$, the relative error in the metric fields and
position of the Lagrange radii is approximately $4\%$, growing to
roughly $15\%$ by the time the event horizon encompasses the entire 
mass distribution.  In all cases, the deviations from the correct
solution take the same form: our computed field values are closer to
unity (and the previous timestep's solution) than we would expect from
the semi-analytic solution. 

This behavior was confirmed by testing the collapse of dust clouds
with initial areal radii of $\bar{r}_s=5$ and $\bar{r}_s=50$.  In both
cases, we find extremely accurate results until the event horizon
forms, at which point our accuracy degrades to a noticeable extent.  The effect
seems to depend primarily on the formation of an event horizon in the
system, and not on the relative change of the density or various field
quantities. We have concluded that the relaxation techniques
used in the code have difficulty in their current form in handling the
steep spatial gradients in the shift function near the event horizon
(see Fig. 2 of \cite{Petrich}),
and we are working on techniques to better handle this situation.
Of particular importance is altering the relaxation parameters of the
iterative scheme 
used by the field solver in the presence of these large field values and
gradients near the event horizon, to correct for the systematic drift
away from the expected values.
\subsection{Circular Orbits of Quasi-equilibrium Configurations}
\label{sec:circ}

Finally, to test out the overall stability of the code, we evolved
several quasi-equilibrium binary configurations using our
code, but without adding in the radiation reaction drag
terms.  Since the CF formalism is time-symmetric, we expect that any
stable quasi-equilibrium model should yield a circular
orbit, maintaining a constant binary separation, ADM mass, and internal
rotation profile, among other parameters.  Here, we chose models at an
initial binary separation of $r_0/{\cal M}_{ch}=19.91$, $20.42$, and
$22.97$ from the $M/R=0.14,~0.14$ equal-mass sequence
of TG, denoting the resulting calculations as runs QC1, QC2, and QC3,
respectively.  For NS with ADM masses of $M_0=1.4~M_\odot$, these separations
correspond to $33.64$, $34.51$, and $38.81~{\rm km}$, respectively.
The innermost of these represented the limiting case found
for this sequence just before formation of a cusp at the inner edge of the
two NS.  

As a first check of our code, we compared the orbital frequency
determined from our dynamical runs to the known value for each model
taken from the quasi-equilibrium sequence.  We found excellent agreement
between the orbital periods computed from our runs, $T/{\cal
  M}_{ch}=422.6$, $438.2$, 
and $506.4$, and those determined by TG, $T/{\cal M}_{ch}=422.8$, $438.0$, and
$519.7$, for runs QC1, QC2, and QC3, respectively.

\begin{figure}
\centering\includegraphics[width=6in]{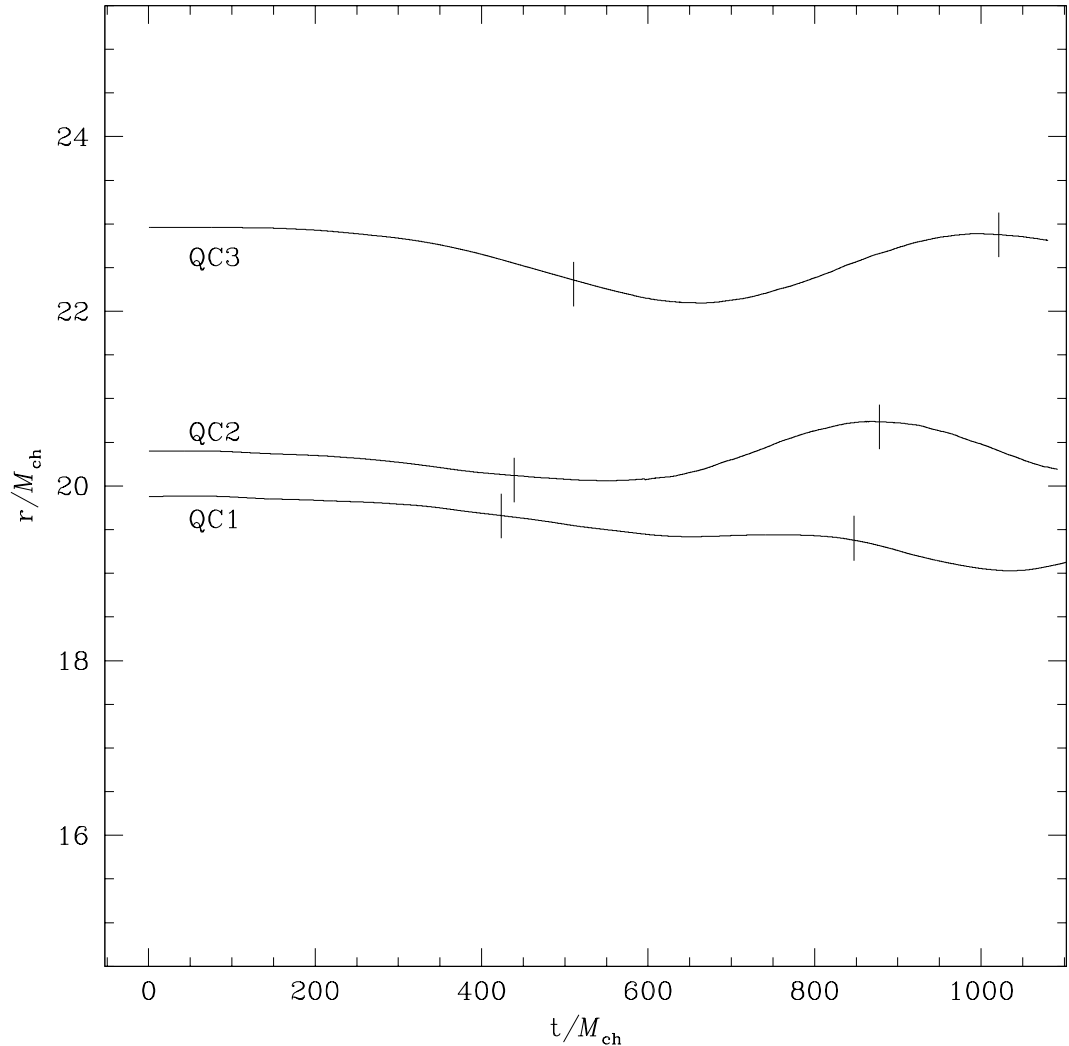}
\caption{Binary separation as a function of time for the binaries in
  runs QC1, QC2, and QC3, using initial data generated by
\protect\cite{TG}, evolved forward in
time with radiation reaction terms neglected.  The calculations use
  binaries with initial separations of 
$r_0/{\cal M}_{ch}=19.91$, $20.42$,
and $22.97$, respectively.  
Full orbital periods, with $T/{\cal M}_{ch}=422.8$, $438.0$, and $519.7$,
respectively, are shown with tick marks.  We see that the resulting
orbits are nearly circular, with changes in separation of no more than
$4\%$ over the first two orbits.  This is the strongest available
evidence that the innermost point along this equilibrium sequence, is
actually stable against merger.} 
\label{fig:circ_rvst}
\end{figure}

There are several conserved quantities which should be respected in
the time-symmetric CF formalism, allowing for further code tests.
In Fig.~\ref{fig:circ_rvst}, we show the evolution of the binary
separation for the runs.  For each run, the orbital period is shown with tick
marks.  We
see in each case that the orbit is stable, with variations in the
binary separation of no more than $4\%$ during the first two orbits.
The timescale for the radial variations is similar to the orbital
period, but not an exact match; this reflects both the effects of GR
as well the slight degree of time-asymmetry present in the numerical
implementation of the CF formalism.  We note that the deviations from
circularity were largest for run QC3, which had the largest binary
separation.  We believe this results from the larger relative
magnitude of the spurious initial velocities that result from
deviations away from equilibrium in the initial condition; while these
terms are of essentially constant magnitude in all three runs, the
equilibrium velocity field has the smallest magnitude at the largest
separation.

Run QC1 was started from the
innermost point along this binary NS equilibrium sequence, and the
binary performs three complete orbits with no sign of plunging behavior. 
AS such, these results can be taken as the first direct proof that the entire
equilibrium sequence is stable, and suggest that these configurations should be
reasonably accurate approximations to the true physical state of
merging binaries.  Further evidence of this claim is presented below
in Sec.~\ref{sec:unstable}, 
when we describe the results of our calculations with
radiation reaction effects included.
Note that this result is not unexpected given the absence of a turning point (minimum of
ADM mass and total angular momentum) along this irrotational equilibrium
sequence. Indeed, while such a turning point along an equilibrium sequence
of {\em corotating\/} binaries marks the onset of {\em secular
instability\/}, a true {\em dynamical instability\/} is usually associated
with a turning point along an {\em irrotational\/} sequence \cite{LRS2,LomRS,TG3}.

While this result may appear at first glance to disagree with
those of /cite{Marro}, who find that the ISCO occurs at a greater
binary separation than the termination point of an equilibrium
sequence, we believe that the difference is purely semantic.  We define an initial
configuration to be dynamically stable if {\em in the absence of
dissipative radiation reaction effects}, the circular orbit remains
stable (no merger occurs) when evolved forward in time.  In \cite{Marro}, a
configuration is described as within the ISCO if in full GR (i.e.,
including dissipative radiation reaction effects), a binary starts
merging (the surfaces of the two NS come into contact) {\em within one orbit\/}
when evolved forward in
time.  Clearly, using these definitions, the same initial
configuration can be dynamically stable (in the time-symmetric CF
sense) at a separation within the ``ISCO'' as defined by \cite{Marro}.

\begin{figure}
\centering\includegraphics[width=6in]{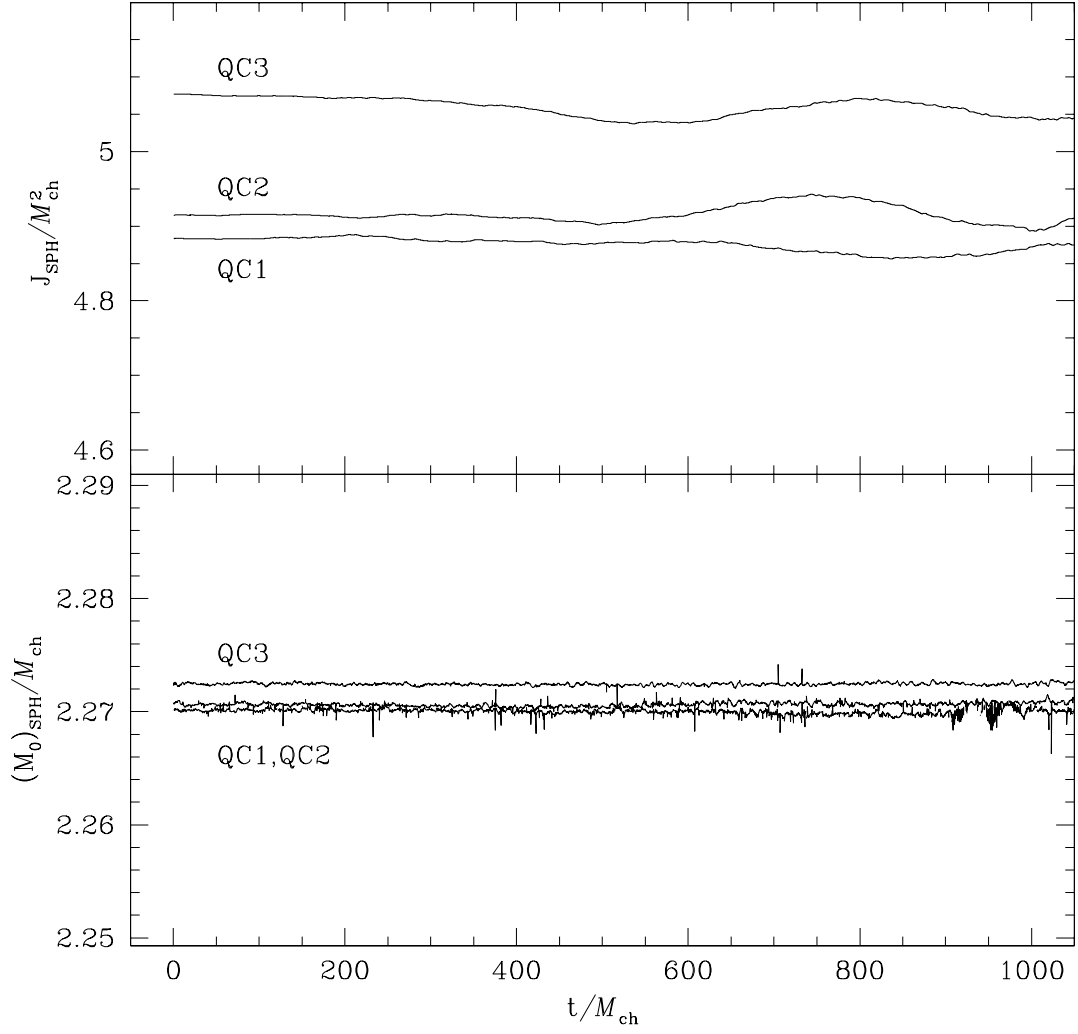}
\caption{ADM mass and total angular momentum, calculated from
Eqs.~\protect\ref{eq:adm_sph}  and \protect\ref{eq:j_sph}, runs QC1,
QC2, and QC3.  We see that the ADM
mass is conserved almost exactly, and the angular momentum to within
$1\%$, with the variation occurring primarily on the orbital timescale.}
\label{fig:circ_mj_sph}
\end{figure}

\begin{figure}
\centering\includegraphics[width=6in]{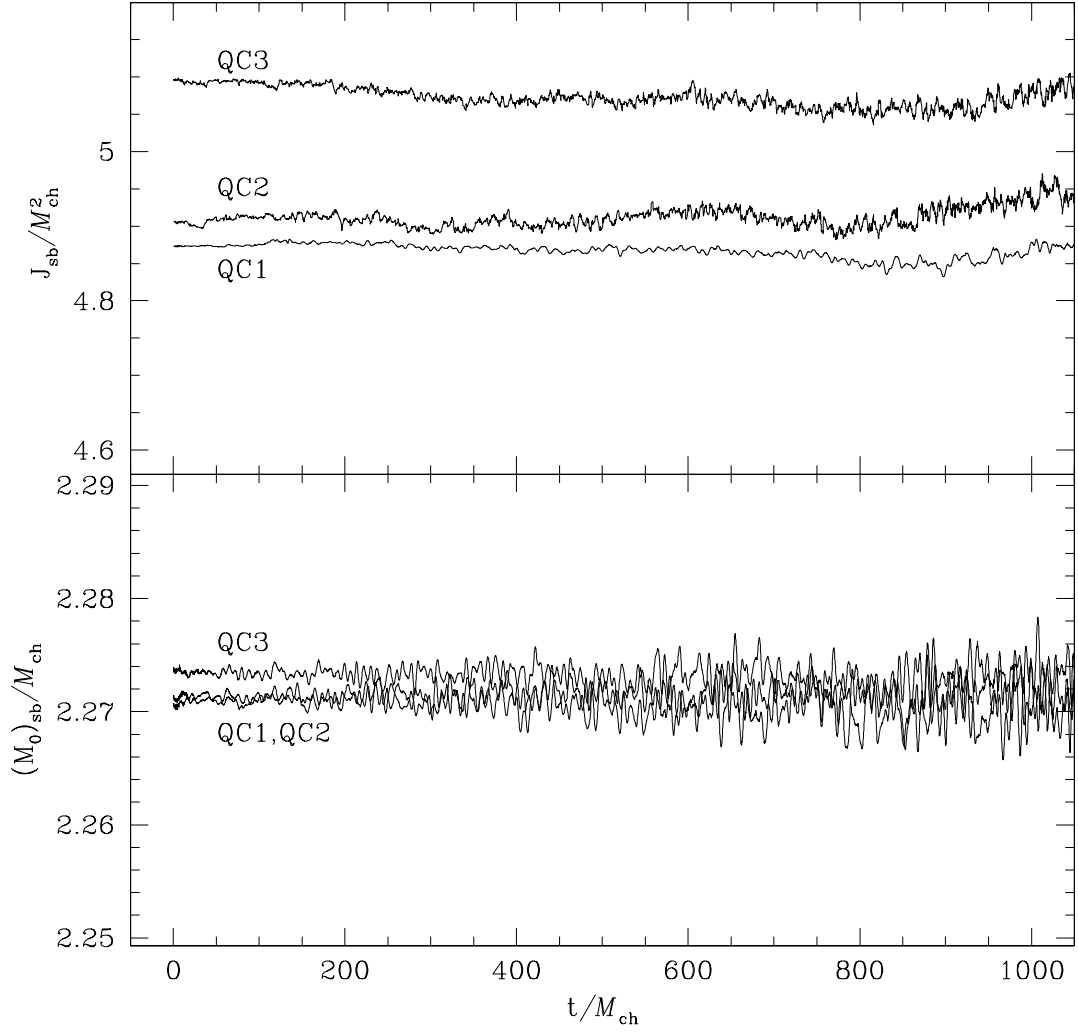}
\caption{ADM mass and total angular momentum, calculated in the
spectral basis from
Eqs.~\protect\ref{eq:adm_sb}  and \protect\ref{eq:j_sb}, for the
runs shown in Fig.~\protect\ref{fig:circ_rvst}.  We see roughly the
same amount of variation in the angular momentum as was found for SPH
summation in Fig.~\protect\ref{fig:circ_mj_sph}.
The ADM mass shows considerably more variation than the SPH version,
but remains well within $1\%$ of the original value with no systematic drift.}
\label{fig:circ_mj_lor}
\end{figure}

In order to estimate how well our
code respects conserved quantities, we show in Figs.~\ref{fig:circ_mj_sph}
and \ref{fig:circ_mj_lor} the evolution of both the ADM mass and system
angular momentum, calculated from the SPH expressions,
Eqs.~\ref{eq:adm_sph} and \ref{eq:j_sph}, and the spectral basis
forms, Eqs.~\ref{eq:adm_sb} and \ref{eq:j_sb}, respectively.
In the former, we see that the SPH expression for the ADM mass is
remarkably constant over time, with only very minor deviations of
relative magnitude less than a tenth of a percent.  The system angular
momentum varies more, but is still conserved to within $1\%$, with no
sign of a systematic drift in either direction.  We note that much of
the variation is correlated with the deviations in the binary
separation, oscillating on the orbital timescale.  We see much more
variation on an iteration to iteration basis when we look at the same
quantities computed using the spectral basis.  This is hardly
surprising, since these values are computed at the end of a 
relaxation routine, and we expect some degree of variation on a
step-by-step basis depending on the exact phase space path traced out by the
iterative solution.  All the same, we see that the angular momentum is
conserved to roughly the same level in the spectral basis as it was
when computed using SPH, and while the variation in the ADM mass is
larger, we see no sign of a systematic drift.

\section{Dynamical Calculations}\label{sec:results}

While dynamical calculations including the effects of radiation
reaction are the only way study the coalescence of binary NS systems,
they have several additional uses which are often overlooked.  Of
particular importance is the ability to determine the validity of
quasi-equilibrium models as initial conditions for dynamical
calculations, regardless of whether CF or full GR gravity is used.
Thus, we computed two dynamical runs including radiation reaction.
Run RR1 was started from the cusp point of the sequence (the same
initial configuration used in run QC1),
and was used to study the details of the coalescence process.
Run RR2 was started from a larger separation (the same initial
configuration used in Run QC3), and was
used to test out the deviations we expect from quasi-equilibrium prior
to reaching the termination point along the equilibrium sequence.

\subsection{Stable Regime}\label{sec:stable}

\begin{figure}
\centering\includegraphics[width=6in]{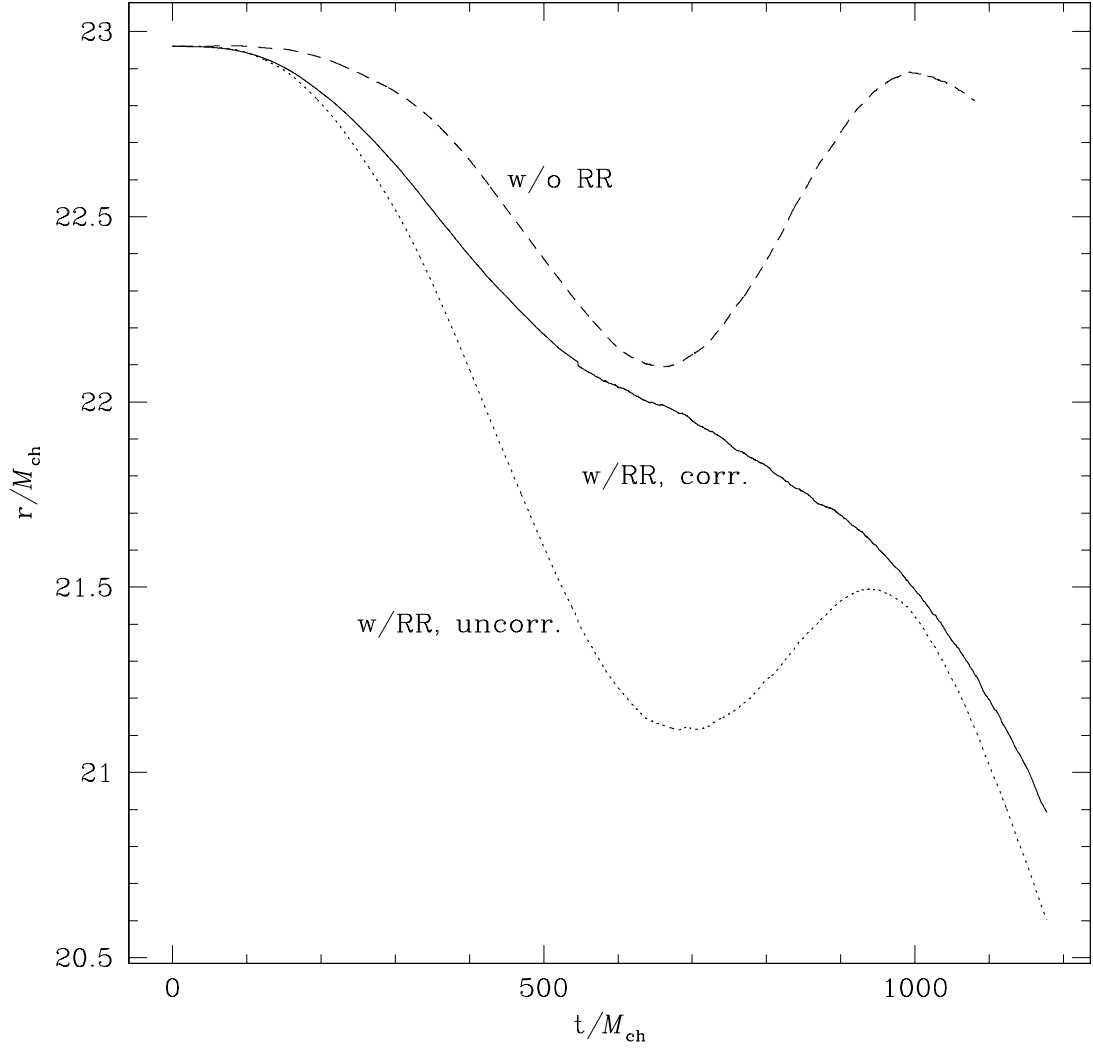}
\caption{Binary separation over time for run RR2, started from an
  initial separation of $r_0/{\cal M}_{ch}=22.97$, the same initial
  configuration that was used for run QC3.  The dotted line shows the
  original, ``uncorrected'' result, including a separation increase
  from $t/{\cal M}_{ch}=700-950$.  This is primarily due to oscillations
  associated with numerical noise and deviations from equilibrium in
  the initial configuration.  Correcting for deviations from
  circularity in run QC3, which ignored radiation reaction (dashed
  curve), yields the ``corrected'' result, shown as a solid line.  We
  see monotonic decrease in the separation over time, with an
  ellipticity induced by our initially circular orbit.}
\label{fig:dyn_rvst}
\end{figure}

The evolution of the binary separation for run RR2
is shown in Fig.~\ref{fig:dyn_rvst}.  The dotted curve, showing
the result from the calculation, does not have the behavior one would
expect.  Notably, after the binary separation decreases monotonically
until reaching $r/{\cal M}_{ch}=21.2$ at $t/{\cal M}_{ch}=700$, 
the system turns around and
expands briefly back to a separation of $r/{\cal M}_{ch}=21.5$ at 
$t/{\cal M}_{ch}=950$ before
shrinking again.  This does not reflect any inherent problem in our radiation
reaction formalism, since oscillations in the binary orbit were found
for calculations which ignored all radiation reaction effects.  In
fact, if we ``correct'' the binary separation by looking at the
difference in separation at equivalent times between runs RR2 and QC3, which were started
from the same initial configuration, 
with and without radiation backreaction terms, we see a
pattern of monotonic decrease, shown as a solid line.  This seems to
indicate that deviations from circularity in the orbit of the
quasi-equilibrium binary configurations represent a systematic effect
in the evolution. 

The ``corrected'' infall curve shows clear signs of an orbital
eccentricity with a timescale roughly corresponding to the orbital
period.  This is a natural consequence of starting out from an initial
condition with zero infall velocity, and has been seen before in
virtually every PN and CF calculation
(FR, \cite{APODR,Oech}).  Its origins are clear: the
framework used by GGTMB to construct quasi-equilibrium initial 
conditions assumes a
helical Killing vector exists, which enforces an initial circularity in the
orbit, rather than the proper infalling trajectory.  
If calculations could be started from sufficiently large separations,
GW emission would cause the orbit to circularize, but the process works
slowly, and breaks down as the binary makes the transition toward
a dynamical merger.  It is fair to say that no computational scheme
can currently be trusted to remain stable over the period required to
circularize the orbit. 

\subsection{Coalescence}\label{sec:unstable}

\begin{figure}
\centering\includegraphics[width=6in]{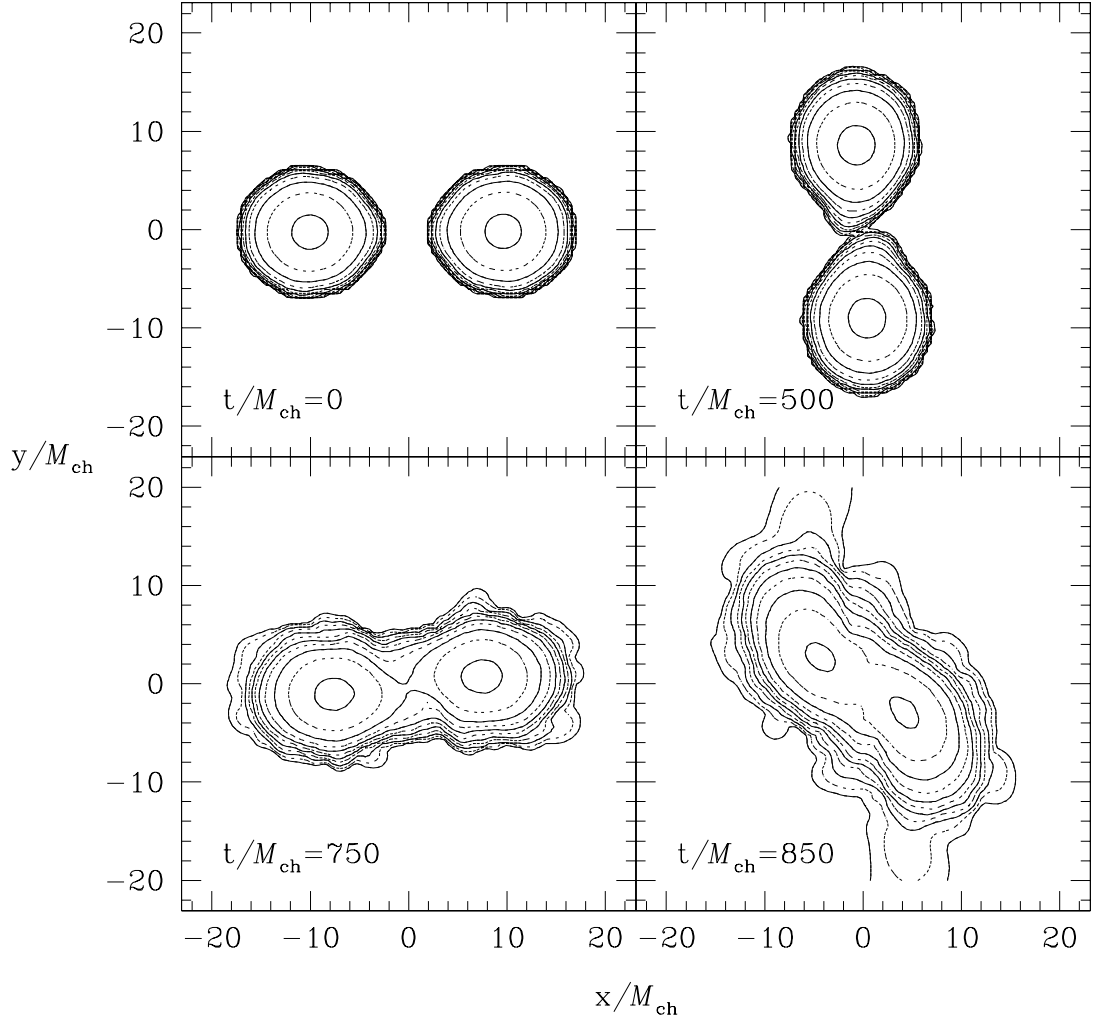}
\caption{The evolution of the matter run RR1, started from
  just outside the separation where a cusp develops.   
  We followed the evolution through the merger and formation of a
  remnant.  Density contours are logarithmically spaced, two per
  decade, ranging from ${\cal M}_{ch}^2\rho_*=10^{-6.5}-10^{-1}$.  We see the
  development of significant tidal lag angles at $t/{\cal
  M}_{ch}=500$, followed by an ``off-center'' collision.  This process  
leads to the formation of a vortex sheet
  and a small amount of matter ejection by $t/{\cal
  M}_{ch}=850$ from matter running along the surface of the other NS.}
\label{fig:rr_dv1}
\end{figure}

We have also computed the full dynamical evolution of a binary system
started from the innermost
point along the quasiequilibrium sequence, denoted run RR1.  In
Fig.~\ref{fig:rr_dv1}, we show density contours from
the system during the
binary phase, which lasted until $t/{\cal M}_{ch}=883$.  The contours are equally
spaced logarithmically, two per decade, ranging from density values of
${\cal M}_{ch}^2 \rho_*=10^{-6.5}-10^{-1}$.  We recognize a familiar pattern
from past PN and relativistic calculations (FR, \cite{APODR,Oech,STU}),
including the development of tidal lag angles as the
timescale on which the gravitational field evolves becomes comparable to
the dynamical timescale.   This gives rise to
an ``off-axis'' collision, as matter from the inner portion of each star
runs along the trailing edge of the other, forming a turbulent vortex
sheet (see FR3).  Some fraction of the mass in these flows eventually
crosses through the outer Lagrange point on the opposite side of the
binary, forming the
very low-density spiral arm structures seen at $t/{\cal M}_{ch}=850$.  
In contrast to Newtonian binaries, in which angular momentum transfer
outward leads to massive spiral arm formation, the very low-density
arms formed here have velocities much smaller than the escape velocity, and the
vast majority of the mass remains gravitationally bound to the
system.

\begin{figure}
\centering\includegraphics[width=6in]{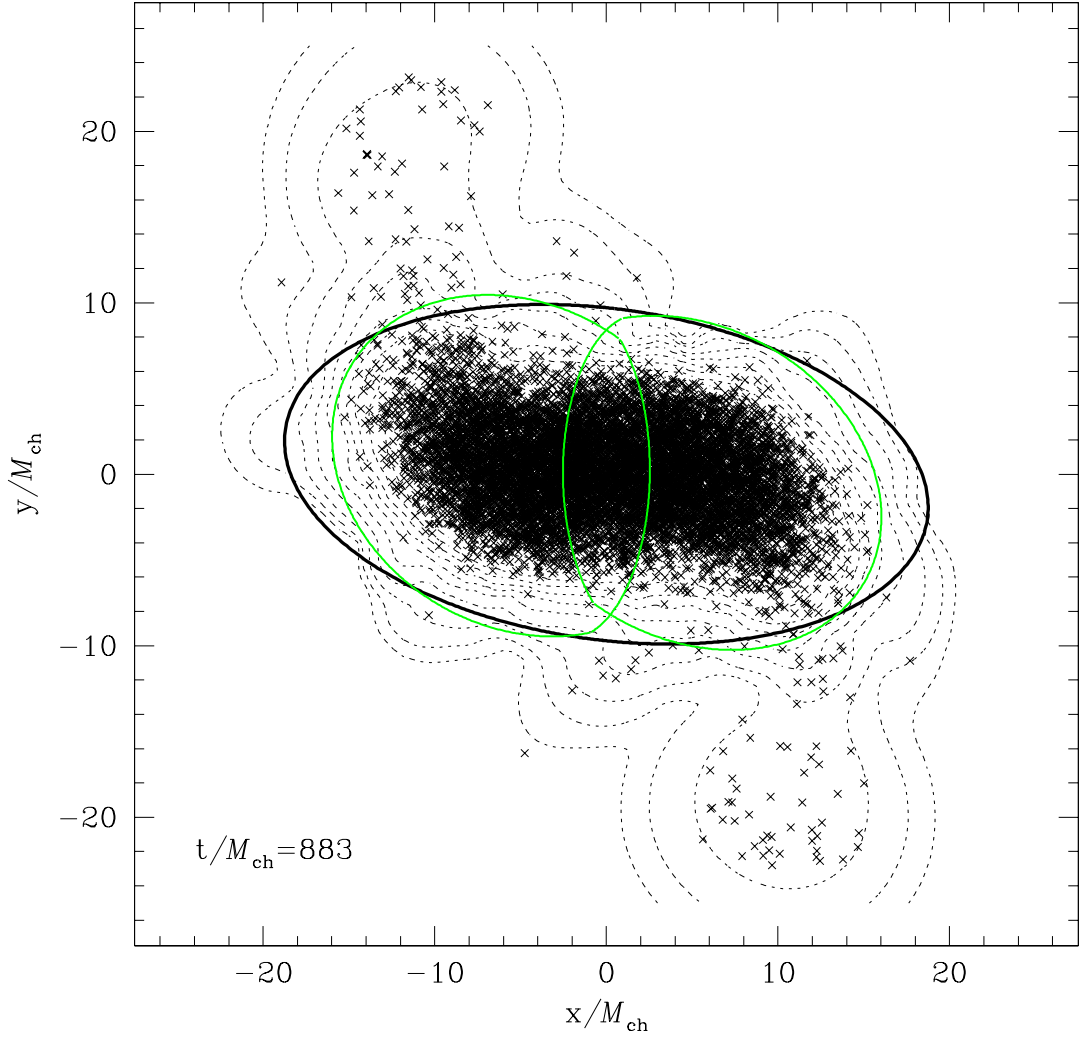}
\caption{Detailed view of the binary to single-star transition performed during
  run RR1 at
  $t/{\cal M}_{ch}=883$.  We show the positions of all SPH particles, as well as
  density contours, shown as dashed lines, logarithmically spaced two
  per decade. The surface of the inner computational domain in both
  the binary and single-star representations are shown as heavy solid
  lines, with good agreement between the two.} 
\label{fig:trans_conv}
\end{figure}

\begin{figure}
\centering\includegraphics[width=6in]{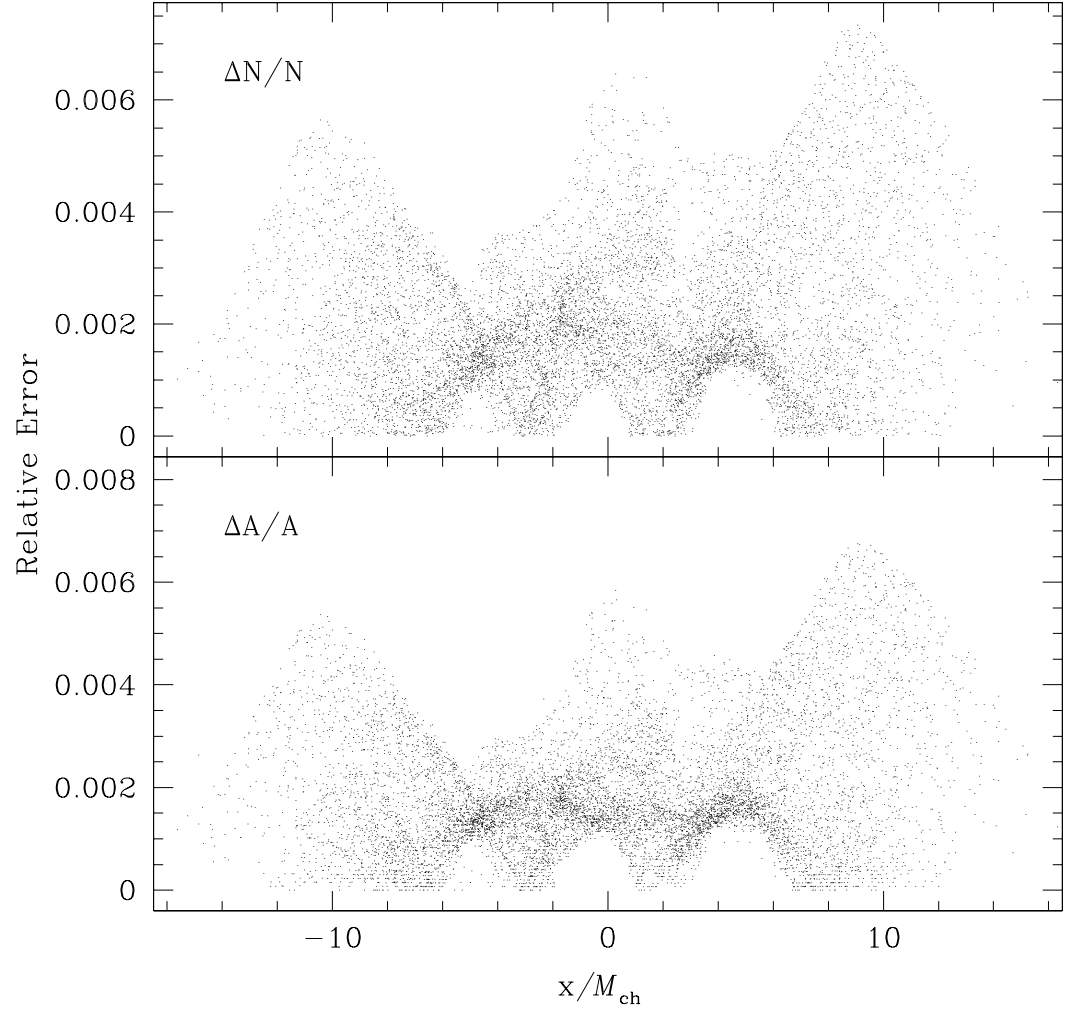}
\caption{The relative change in the lapse function $N$ (top panel) and
  the conformal factor $A$ immediately preceding and following the
  conversion from binary to single-star representations during run RR1
  for every
  particle, shown as a function of the particle's position in the
  x-direction.  The maximum error is approximately $0.8\%$, with a
  mean difference of $\sim 0.2\%$.} 
\label{fig:trans_an}
\end{figure}

At $t/{\cal M}_{ch}=883$, shortly before the binary field solver fails to
converge, we take our matter and field
configurations and transform them into the single-star description
described in Sec.~\ref{sec:merging}.  In Fig.~\ref{fig:trans_conv}, we
show the particle configuration at $t/{\cal M}_{ch}=883$, 
which can be described as
a bar with two low-density arms trailing off the edges.  While it may
seem inappropriate at first to describe the configuration as an
ellipse, we note that the low-density contours, shown as dashed lines,
do form a much more elliptical pattern than one might at first expect.
In interpreting SPH particle plots, it is important to remember that
low-density particles have
large smoothing lengths, implying that the matter distribution extends
well beyond the apparent sharp edge.  The boundaries of the innermost
computational domains are shown in the figure as heavy solid lines,
for both binary and single-star configurations.  We see that they
align rather well, failing to overlap only in low-density regions near
the boundary.  To confirm the validity of the switch, we compare
the field solutions for the particles before and after the
transition.  In Fig.~\ref{fig:trans_an}, we show the relative change
in the lapse  function $N$ (top panel) and conformal factor $A$
(bottom panel), as a function of the x-coordinate.  We see that in all
cases the relative error is $<1\%$.

\begin{figure}
\centering\includegraphics[width=6in]{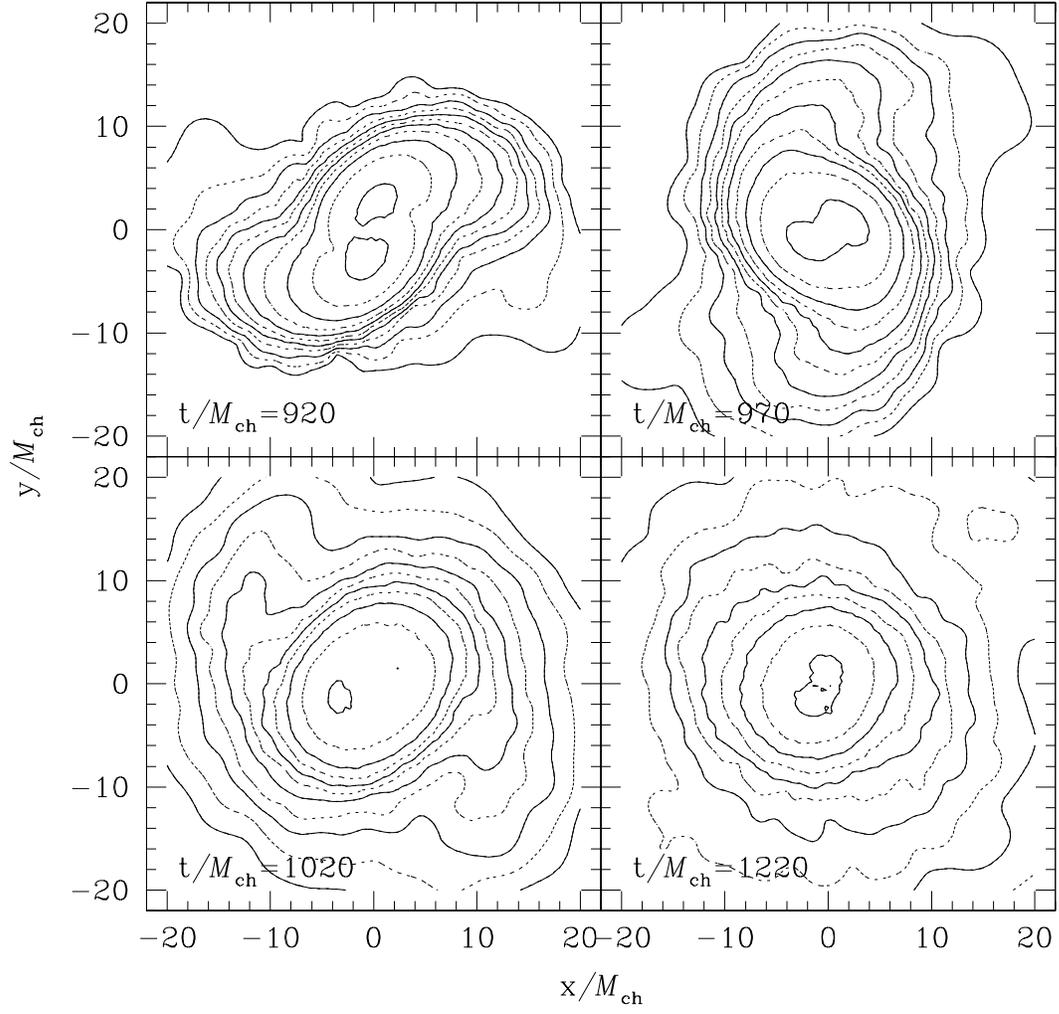}
\caption{The evolution of the matter in run RR1, 
after the transition to a single-star
  representation, following the same conventions
  Fig.~\protect\ref{fig:rr_dv1}.  We
  see that a dense remnant forms in the center of the system,
  surrounded by a thin halo.  Some ellipticity is seen shortly after
  the merger, but the system quickly relaxes toward a spheroidal
  configuration, with maximum density in the center of the system,
  unlike the toroidal configuration found by \protect\cite{STU}.}
\label{fig:rr_dv2}
\end{figure}

\begin{figure}
\centering\includegraphics[width=6in]{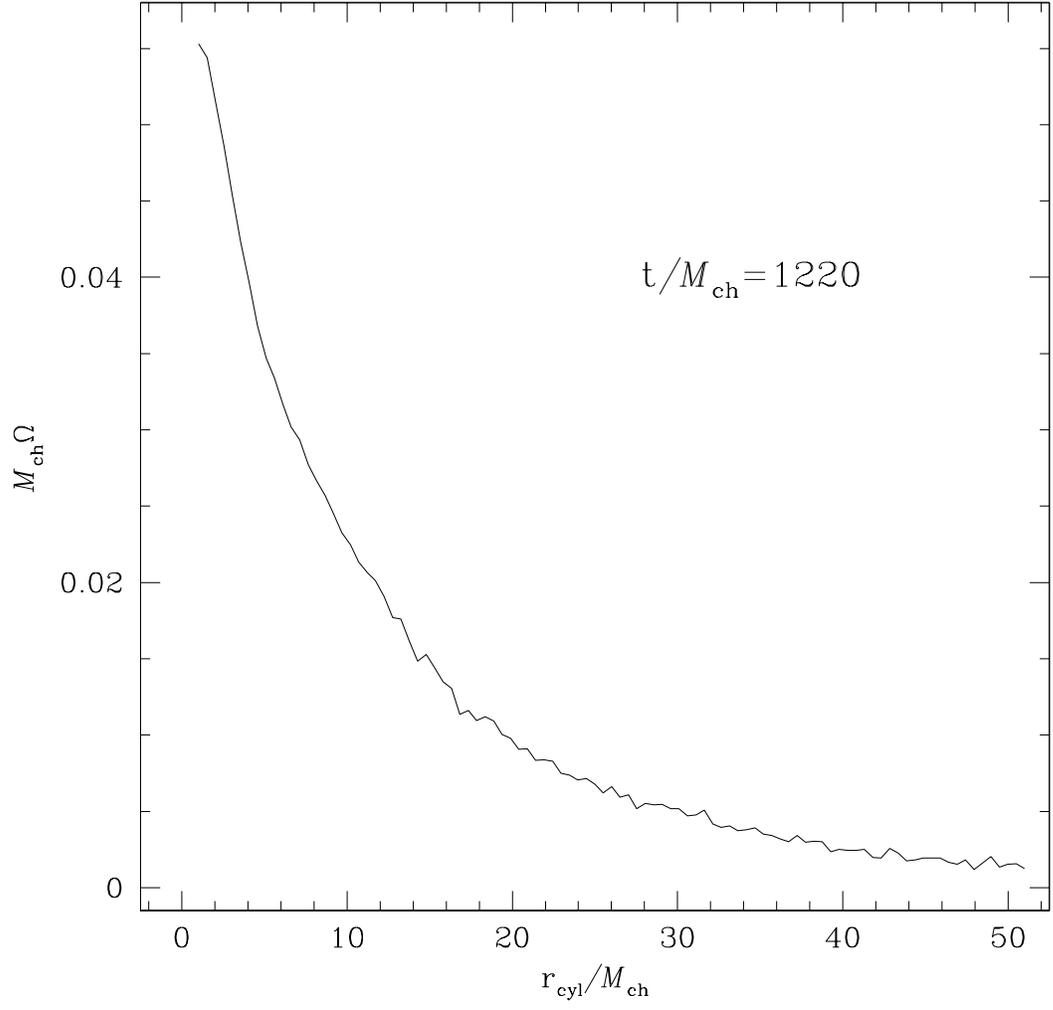}
\caption{Angular velocity of the merger remnant of run RR1 at $t/{\cal
    M}_{ch}=1220$, shown as a
  function of cylindrical radius, $r_{cyl}\equiv \sqrt{x^2+y^2}$.  We
  see strong differential rotation, with the highest angular velocity
  in the center, decreasing monotonically with radius.}  
\label{fig:rr_finvel}
\end{figure}

\begin{figure}
\centering\includegraphics[width=6in]{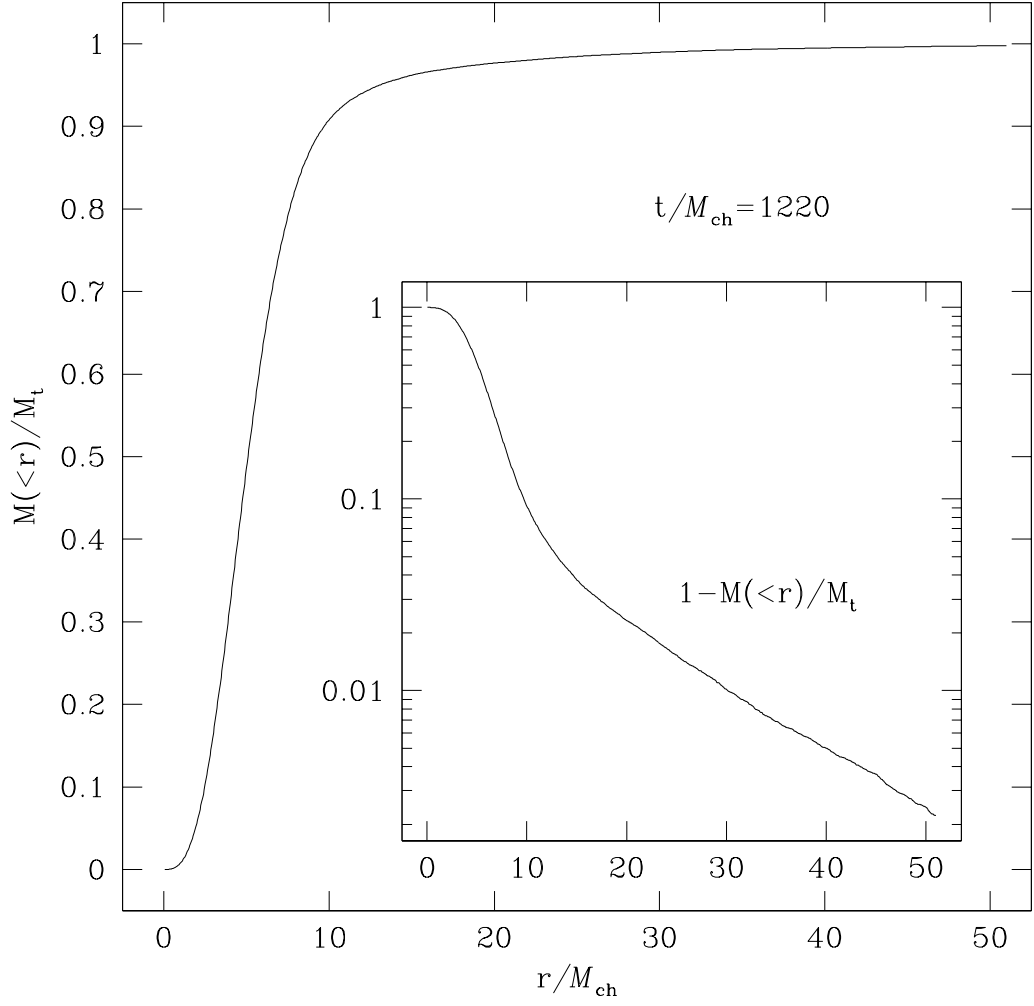}
\caption{Enclosed mass as a fraction of the {\it total} mass of 
the merger remnant in run RR1, at $t/{\cal M}_{ch}=1220$, expressed as a
function of (spherical) radius.  All but a few percent of the total mass
of the system forms the body of the merger remnant, with no more than
a small fraction of a percent ejected from the system.}  
\label{fig:rr_finmass}
\end{figure}

The evolution of the single-body configuration is shown in
Fig.~\ref{fig:rr_dv2}.  We see a strong pattern of differential
rotation in the merger remnant, which slowly relaxes from a very
elliptical shape toward a more spheroidal one.  This is to be
expected, as our choice of EOS with $\Gamma=2.0$ should not be able
to support a long term ellipsoidal deformation \cite{RS1}.  By $t/{\it
  M}_{ch}=1220$, the remnant
has relaxed to a nearly circular profile, but the differential rotation,
as shown in Fig.~\ref{fig:rr_finvel}, persists. 
The latter will dissipate slowly
on either the viscous or magnetic braking timescales, 
beyond the scope of what we can reasonably
calculate \cite{Shap1,CSS,TS}.  
Differential rotation is expected to stabilize the
star against gravitational collapse in the short-term \cite{BSS,LBS}.  
Quantitatively
accurate determinations of the rotational velocity profile in terms of
the parameters of the initial system are likely to be crucial for
making prediction as to which systems will or will not collapse
promptly to BHs, especially since these systems will likely have very
large masses.  As we see in Fig.~\ref{fig:rr_finmass}, the vast
majority of the rest mass of the system ends up in the merger remnant
itself, with a fraction of a percent of the total mass forming a
low-density, bound halo around the remnant.  This seems to be the
consensus from PN and relativistic calculations of
irrotational binaries (FR3,\cite{STU}) and even PN and relativistic
calculations of synchronized binaries (FR3,\cite{Oech}), which
traditionally yielded significantly higher mass ejection fractions in
Newtonian calculations.  We note that PN calculations of
irrotational binaries yielded an ejected mass fraction of $\lesssim
1\%$ (FR3), rather than the $6\%$ quoted by \cite{STU}.

\begin{figure}
\centering\includegraphics[width=6in]{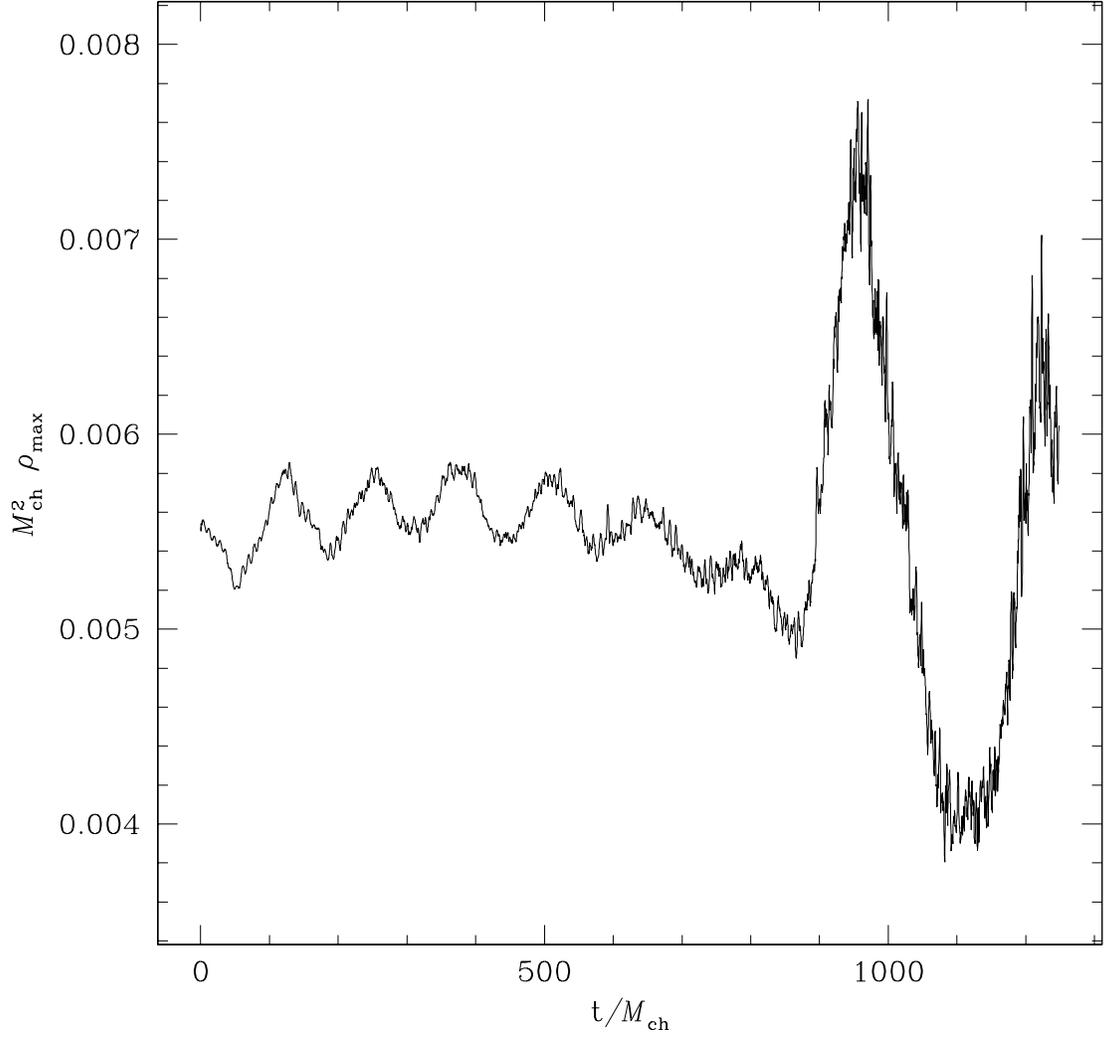}
\caption{Maximum density as a function of time for run RR1.  
We see the SPH configuration
  oscillates slightly around equilibrium, decreasing slowly as the
  binary plunges toward merger.  During the merger, we see a sharp
  decrease, followed by a large spike upward and evidence for sharp,
  non-sinusoidal oscillations.}
\label{fig:rr_rho}
\end{figure}

To compare with the relativistic results of \cite{Oech,STU}, we show the
evolution of the maximum density as a function of time in
Fig.~\ref{fig:rr_rho}.  We see at the earliest times a low-amplitude
pulsation, resulting 
from small deviations from equilibrium in the initial SPH particle
configuration.  This pulsation
damps away almost completely by the time the stars plunge inward.  As the
system begins to accelerate rapidly inward prior to the
merger itself, the maximum density decreases as the stars are tidally
stretched.  This effect, seen in a number of calculations, further
indicates that  these systems will not undergo a pre-merger
gravitational collapse.
This process was originally suggested in \cite{WMM}, but is likely
to have been a numerical artifact since they used a version
of the CF formalism containing an error in one of the evolution
equations.  Indeed, the quasi-equilibrium sequences in TG show a slow
{\it decrease} in the central baryonic and energy densities, $\rho_*$
and $\rho$, as the binary separation decreases.
After contact, we see a strong rise in the maximum
density, followed by a rapid, high-amplitude oscillation.  This result
is similar to that seen by \cite{Oech}, although they found a
relatively higher maximum density value during both the peak and the trough
of the oscillation. This is almost certainly a result of using
different initial spin configurations.  Irrotational systems, such as
the one used in run RR1, concentrate relatively more angular momentum
at small radii compared to initially synchronized systems, like run A
of \cite{Oech}.  Thus, when a remnant is formed from an initially
irrotational binary, the central density
will typically be lower, since there is a greater centrifugal barrier
and less pressure support is needed to stabilize the configuration.
Our results differ rather significantly from those of \cite{STU}, who
found a smaller-amplitude, more sinusoidal oscillation after merger.
It is possible that this discrepancy can be attributed to the use of
the CF approximation rather than full GR.  A much more likely
explanation, however, is that the difference results from the
numerical methods
used to describe shock heating in the matter.  
Lagrangian SPH codes were used here and in
\cite{Oech}, whereas Shibata and collaborators \cite{SU1,SU2,STU}
use an Eulerian grid-based code, which may have a better
ability to resolve shock fronts.  It seems clear by examining the
results from run M1414
in \cite{STU} that as the NS cores
converge, the increase in the central density is suppressed by the
conversion of kinetic energy into heat.  The merger remnant
shows a spike in the internal energy in the very center, $2-3$ times
the adiabatic value, and a corresponding decrease in the density,
giving the remnant its toroidal profile.  Only calculations using their
new shock-capturing scheme \cite{Shi1} produce this behavior; previous
calculations yielded remnants with centrally condensed density
profiles \cite{SU1,SU2}.  It will be of great future interest to
determine whether or not SPH studies of rapidly rotating collapsing
matter configurations can produce these toroidal configurations with
``hot'' cores, and if so, which aspects of shock physics are crucial
for understanding this process.

\begin{figure}
\centering\includegraphics[width=6in]{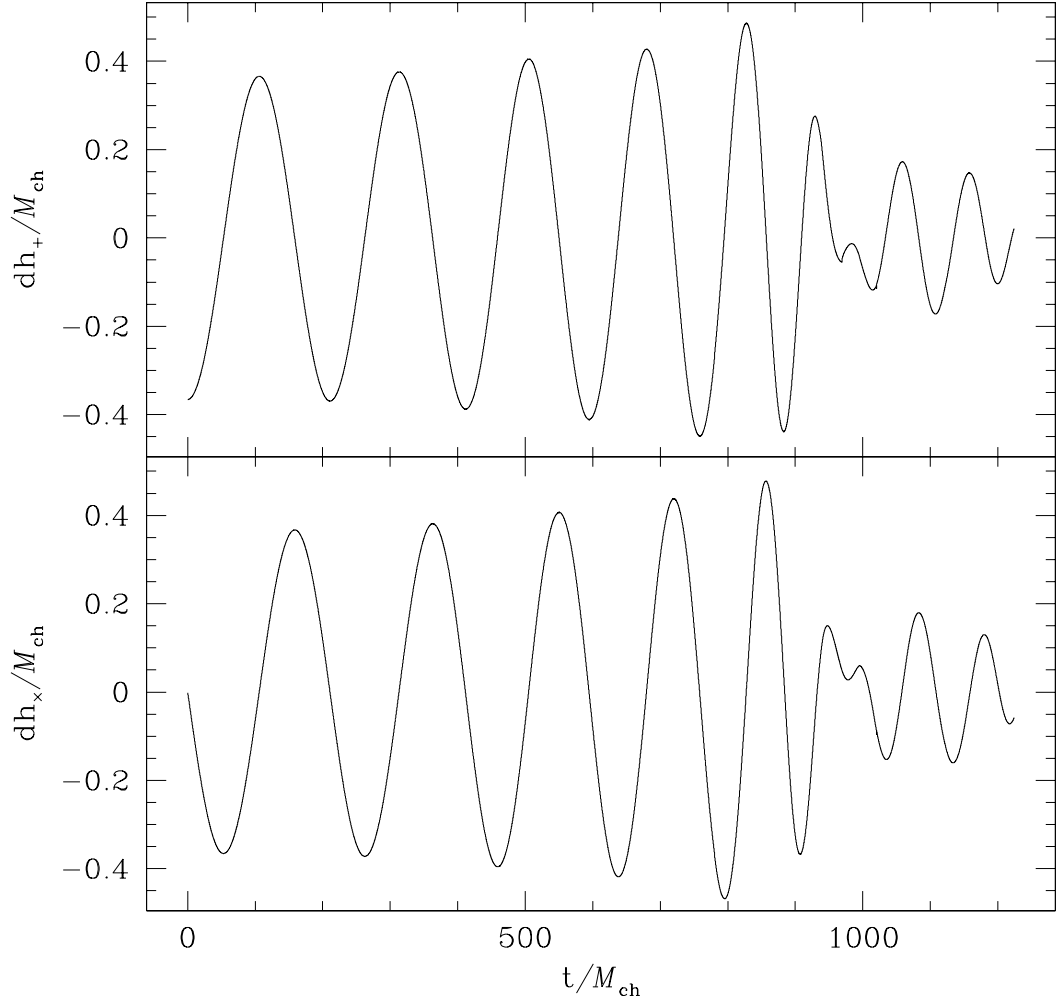}
\caption{The GW signal in the $h_+$ and $h_\times$ polarizations
for run RR1, as seen by an observer situated along
  the vertical axis, following Eqs.~\protect\ref{eq:hplus} and
  \protect\ref{eq:htimes}.  We see a chirp signal followed by a
  modulated ringdown spike.  The modulation is caused by the alignment
  between quadrupole deformations in the inner regions of the remnant
  core and and those at larger radius.  When there is strong
  misalignment, there is destructive interference and the signal
  amplitude drops, as we see at $t/{\cal M}_{ch}=920$ and $t/{\cal
    M}_{ch}=1220$ in Fig.~\protect\ref{fig:rr_dv2}.
.  At $t/{\cal M}_{ch}=1080$ the
  density contours are more aligned, and the amplitude reaches a
  temporary maximum.}
\label{fig:rr_gwpl}
\end{figure}

In Fig.~\ref{fig:rr_gwpl}, we show the GW signal for
run RR1, calculated from Eqs.~\ref{eq:hplus}
and \ref{eq:htimes}.  The waveforms show a
familiar chirp signal up until $t/{\cal M}_{ch}\approx
850$, followed by a modulated, high-frequency ringdown component.
The strength of our signal at peak amplitude matches
extremely well with the results of \cite{Oech,STU}, as one would
expect from simple dimensional analysis.  Our modulated remnant
signal, though, is much more similar to the results of \cite{STU}
than \cite{Oech}, who find a damped ringdown signal of lower
amplitude for this model, with no obvious modulation.  
Previous PN calculations (FR2, FR3) identified the source of the GW
amplitude modulation as a combination of differential rotation and
ellipticity in the remnant.  When the inner and outer regions of the
remnant have ellipsoidal deformations which are roughly aligned, 
as we see at $t/{\cal M}_{ch}\approx
900$ in Fig.~\ref{fig:rr_dv1}, the GW amplitude will be at a
maximum.  When differential rotation drives the inner regions into
misalignment, as we see at $t/{\cal M}_{ch}\approx 1020$ in the same
figure, the GW amplitude reaches a minimum.  Eventually, these
effects dissipate away as the remnant relaxes toward a more
spheroidal configuration.  
We believe the lack of modulation in the waveforms shown in Fig.~10 of
\cite{Oech} merely reflects the use of a different initial spin
configuration.  As stated above, synchronized initial configurations
contain relatively less angular momentum at smaller radii, and the
lack of a centrifugal barrier allows the remnant ellipticity to damp
away much more quickly as the densest regions in the NS cores fall
into the remnant's center.  Our results are in broad agreement with
those of \cite{STU}, who find a modulated waveform immediately after
the merger on a similar timescale as our results,
before a longer-lived, smaller-amplitude damped modulation
eventually appears.

Calculating the GW energy spectrum for the merger waveform is more
complicated than simply applying Eq.~\ref{eq:dedf}, since taking the
Fourier transform of a signal with non-zero initial and final values
introduces aliasing of the boundary conditions into the resulting
waveform.  To correctly derive the proper spectrum, one must also
``attach'' analytic solutions 
to the beginning and end of the calculated signal, representing the
portions of the inspiral and ringdown phases, respectively, which fall
outside the bounds of the numerical evolution.  In the past, the
authors (FR3), and others \cite{ZCM1,ZCM2,Oech}, have modeled the
initial inspiral phase via the Newtonian point-mass approximation, but
that is clearly not physically realistic.  Indeed, it has been shown
in FGRT
that relativistic effects should have a significant effect on the GW
spectrum even before the dynamical merger.
Summarized, we know that for an equal-mass binary
system, the total mass-energy is given by
$E=M_t-M_t^2/8r^2=2^{1.2}{\cal M}_{ch}-{\cal M}_{ch}^2/(2^{0.6}r^2)$, and the
(Keplerian) orbital frequency by $f_{Kep}=(\sqrt{M_t/r^3})/2\pi$.  In
terms of the GW emission frequency, $f_{GW}\equiv
2f_{Kep}=(2^{0.6}/\pi)\sqrt{{\cal M}_{ch}/r^3}$, we find
\begin{eqnarray}
E_N(f_{GW})&=&2^{1.2}{\cal M}_{ch}-\frac{\pi^{2/3}}{2}{\cal
  M}_{ch}^{5/3}f_{GW}^{2/3}, \label{eq:efnewt}\\
\frac{dE_N}{df_{GW}}&=&\frac{\pi^{2/3}}{3}{\cal
  M}_{ch}^{5/3}f_{GW}^{-1/3},\label{eq:dedfnewt}
\end{eqnarray}
where the latter equation demonstrates the familiar power-law
dependence of the GW energy spectrum.

For the quasi-equilibrium sequence from which we take our initial
condition, we found that the system ADM mass can be given in terms of
the GW frequency by a (phenomenological) fit of the form
\begin{equation}
E(f_{GW})/{\cal M}_{ch}=E_N/{\cal M}_{ch}-0.4905({\cal M}_{ch}f_{GW})
+231({\cal M}_{ch}f_{GW})^2. \label{eq:efqe}
\end{equation}
Differentiating this equation yields what we term the
``quasi-equilibrium energy spectrum'', but we require additional
assumptions to be made before we can construct the time-history of the
inspiral waveform.  First, we determine a fit for the GW frequency as
a function of conformal separation, finding that we can approximate
the proper function to within $0.1\%$ with the form
\begin{equation}
{\cal M}_{ch}f_{GW}=\frac{2^{0.6}}{\pi}(r/{\cal
  M}_{ch})^{-1.5}-1.592(r/{\cal M}_{ch})^{-2.5}+6.325(r/{\cal
  M}_{ch})^{-3.5}.\label{eq:frqe}
\end{equation}
Next, we assume that the GW signal amplitude is given by a slightly modified
version of the quadrupole form,
\begin{eqnarray}
Q^{[2]}_{xx}(t)=-Q^{[2]}_{yy}(t)&=&(1.0-\kappa) 2^{0.2}\pi^2{\cal
  M}_{ch}r^2f_{GW}^2 \cos(\theta_{GW}(t)),\label{eq:q2xxqe}\\
Q^{[2]}_{xy}(t)&=&(1.0-\kappa) 2^{0.2}\pi^2{\cal
  M}_{ch}r^2f_{GW}^2 \sin(\theta_{GW}(t)),\label{eq:q2xyqe}
\end{eqnarray}
where the prefactor $\kappa$ is used to match the amplitude of the
inspiral signal onto that at the beginning of our calculated signal.
We find that $\kappa=0.015$ throughout the early phases of our
calculated waveform, indicating that it can be
well-approximated by the expected quadrupole form.  Furthermore, we
note that the resulting energy spectrum is essentially independent of
$\kappa$; as $\kappa$ increases, the GW amplitude decreases,
decreasing both the energy loss rate and the frequency sweep rate in
the same proportion, leaving $dE/df$ unchanged.  Finally, we assume
that the energy loss rate in GWs is given by the standard quadrupole 
expression,
\begin{equation}
dE/dt=0.2\left< 4\pi^2 f_{GW}^2 Q^{[2]}_{ij} Q^{[2]}_{ij}.
\right>\label{eq:dedtqe} 
\end{equation}
To construct the inspiral
waveform, we start from a point along our calculated waveform and
evolve backward in time from that point.  At every timestep, we
calculate the instantaneous energy loss rate from
Eq.~\ref{eq:dedtqe}.  After adjusting the total energy, we calculate
the new GW frequency by implicitly solving Eq.~\ref{eq:efqe}, and
adjust the phase of the GW signal appropriately.  We find the new
binary separation by solving Eq.~\ref{eq:frqe} implicitly, and finally
evaluate the waveform via Eqs.~\ref{eq:q2xxqe} and \ref{eq:q2xyqe}.

The question of where to match the quasi-equilibrium waveform to the
calculated one deserves some attention.  Matching the two at $t=0$ is
very much a mistake, because it represents a transition from an
infalling configuration to a circular one; beforehand, the frequency
sweep rate $df_{GW}/dt$ is positive and increasing, while afterward it
is reset instantaneously to zero.  This mismatch in the infall
velocity, and thus the frequency sweep rate, results in energy
``piling up'' at the transition frequency, as can be seen in FR3 and
to a smaller degree in Fig.~12 of \cite{Oech}.  We find that matching
the inspiral waveform to the calculated signal at $t=0$ reproduces
this error, but that by $t/{\cal M}_{ch}=250$, the match in the
inspiral velocity is sufficient to leave no measurable trace in the
resulting spectrum.

We note that \cite{Oech} also match their inspiral waveform to their
calculated one at some time into the calculation, but we believe that
they place too much trust in the behavior of the energy spectrum near
this transition frequency.
In their paper, they alter the
frequency of a Newtonian inspiral waveform to match their relativistic
calculation by adjusting by hand the coalescence time, $\tau\equiv
(dr/dt)/r$.  We believe this approach to be a mistake.  The effect of this
change is to use a Newtonian waveform with different physical
parameters from those used in the calculation; in particular, it is
equivalent to calculating a Newtonian waveform using the wrong {\it total
mass}, and thus the wrong limit for the spectrum at low frequencies.
It is only through the use of an inspiral waveform whose frequency
approaches the proper Keplerian limit at low frequencies and the {\it
  relativistic} form at high frequencies that accurate energy spectra
can be constructed.  Indeed, since our initial condition was taken
from the same quasi-equilibrium sequence used to generate our
frequency data, the initial GW frequency derived from our calculation
matched that of the inspiral waveform within $0.1\%$ without requiring
further manipulation.

\begin{figure}
\centering\includegraphics[width=6in]{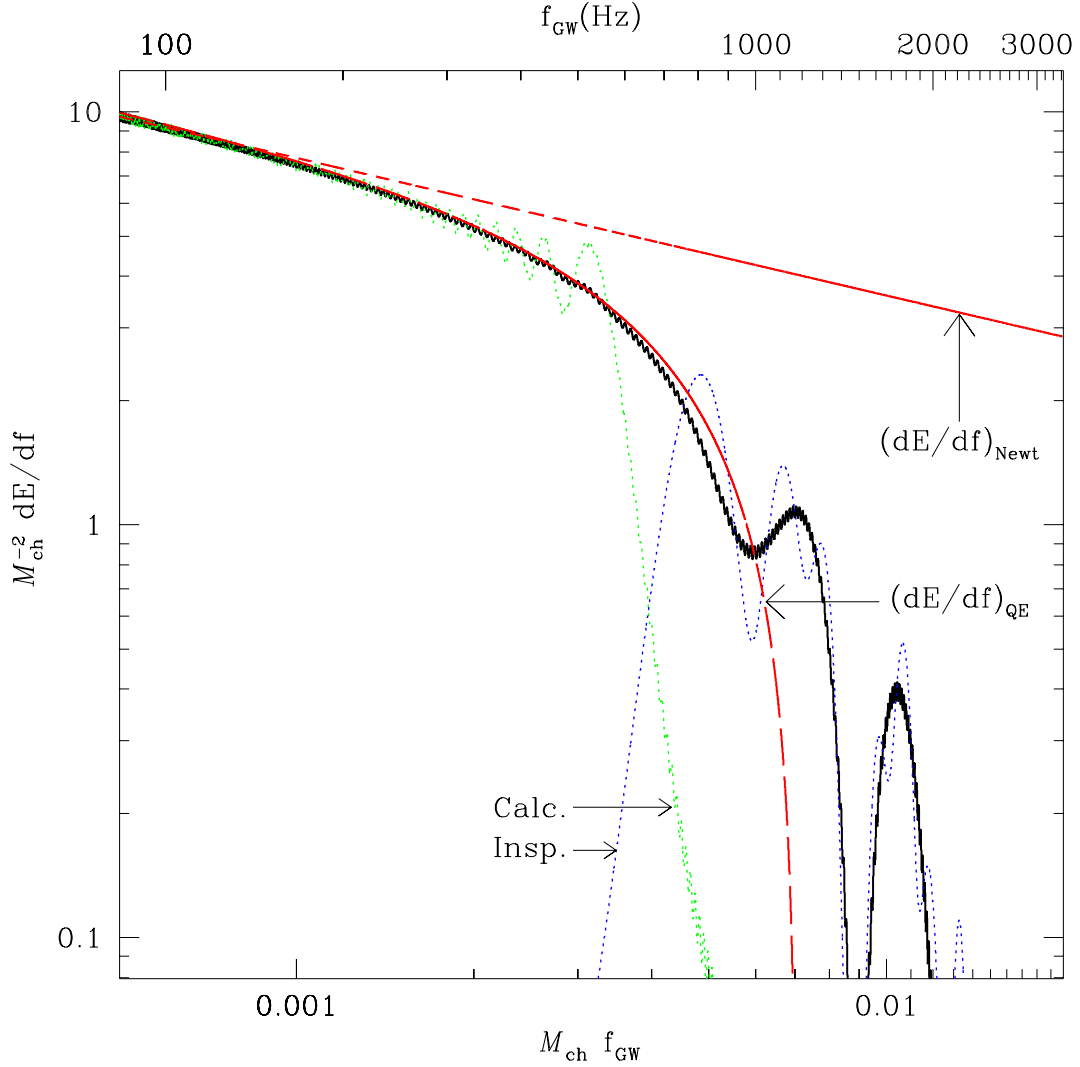}
\caption{GW energy spectrum, ${\cal M}_{ch}^{-2}dE/df$, as a
  function of the GW frequency, ${\cal M}_{ch} f_{GW}$, for run
  RR1. The dotted lines show, respectively at high and low
  frequencies, the components contributed by our calculated signal
  and the quasi-equilibrium inspiral component.  Also shown are the Newtonian
  point-mass energy spectrum $(dE/df_{GW})_{Newt}$ 
(short-dashed line), Eq.~\protect\ref{eq:dedfnewt},
  and the quasiequilibrium fit $(dE/df_{GW})_{QE}$ derived from
  Eq.~\protect\ref{eq:efqe}.   We see
  confirmation that the ``break frequency'' calculated from a fit of
  $E(f)$ for the equilibrium sequence (i.e., the frequency at which
  the energy spectrum decreases  to a given fraction of the Newtonian
  level) is reproduced by a full numerical evolution.  On the
  upper axis, we show the corresponding frequencies in Hz assuming the
  NS each have a mass $M_0=1.4M_\odot$ The two peaks correspond to
  the phases of maximum GW luminosity and ringdown oscillations, respectively.}
\label{fig:rr_gwps}
\end{figure}

In Fig.~\ref{fig:rr_gwps}, we show as a solid line what we believe to
be the first complete and consistent relativistic waveform for
a binary NS merger at frequencies $f_{GW}\le
1.5$ kHz.  The frequencies listed on the upper axis assume our
``standard model'' parameters, i.e., each NS has an ADM mass $M_0=1.4
M_{\odot}$.  The two dotted lines show the components which make up
the energy spectrum.  At low frequencies, ${\cal M}_{ch}f_{GW}<
0.004$,  we see the primary contribution is from the quasi-equilibrium inspiral
waveform, whereas at higher frequencies it is from our calculated
waveform.  The short-dashed line shown the Newtonian point-mass
relation, given by Eq.~\ref{eq:dedfnewt}, and the long-dashed curve
the quasi-equilibrium result,found by numerically differentiating
Eq.~\ref{eq:efqe}.  We see excellent agreement between our calculated
waveform and the quasi-equilibrium fit, up until frequencies ${\cal
  M}_{ch}f_{GW}\approx 0.007-0.009$.  This peak represents the
``piling up'' of energy at the frequency corresponding to the phase of
maximum GW luminosity, as the stars make contact and the infall rate
drops dramatically.  The second peak, at ${\cal
  M}_{ch}f_{GW}\approx 0.010-0.011$, represents emission from the
ringdown of the merger remnant.  It is likely that we underestimate
the true height of this second peak, since we assume that the GW
signal after our calculation damps away exponentially.  Still, it is
extremely unlikely that including the ringdown phase will increase the
strength of this peak by more than a factor of a few, since our chosen
EOS will not support a long-lived ellipsoidal deformation, and it is
likely that the ringdown oscillations will smear the GW emission over
a small range of frequencies rather than coherently emitting at a
single frequency.  

In general, the energy spectrum we calculated here confirms the general
conclusions we put forward in FR3 and FGRT, 
albeit in a much more consistent way.  The GW energy spectrum does show
a significant drop away from the Newtonian point-mass form at
frequencies significantly below 1 kHz, in almost the exact same form
as we predicted from quasi-equilibrium data alone in FGRT.  
Nowhere does the spectrum rise
above the Newtonian value, including the peaks associated with maximum
GW luminosity and the ringdown oscillations.  These results
suggest that the weak signal amplitude of the peaks above 1 kHz,
which lie outside the Advanced LIGO broadband frequency range,
may inhibit detections by high-frequency narrow-band interferometers
as well.  However, combining lower-frequency narrow-band detectors
with broadband LIGO measurements, as suggested in \cite{Hughes},
appears extremely feasible, and may 
allow GW measurements to constrain the NS compactness and EOS.  

\begin{figure}
\centering\includegraphics[width=6in]{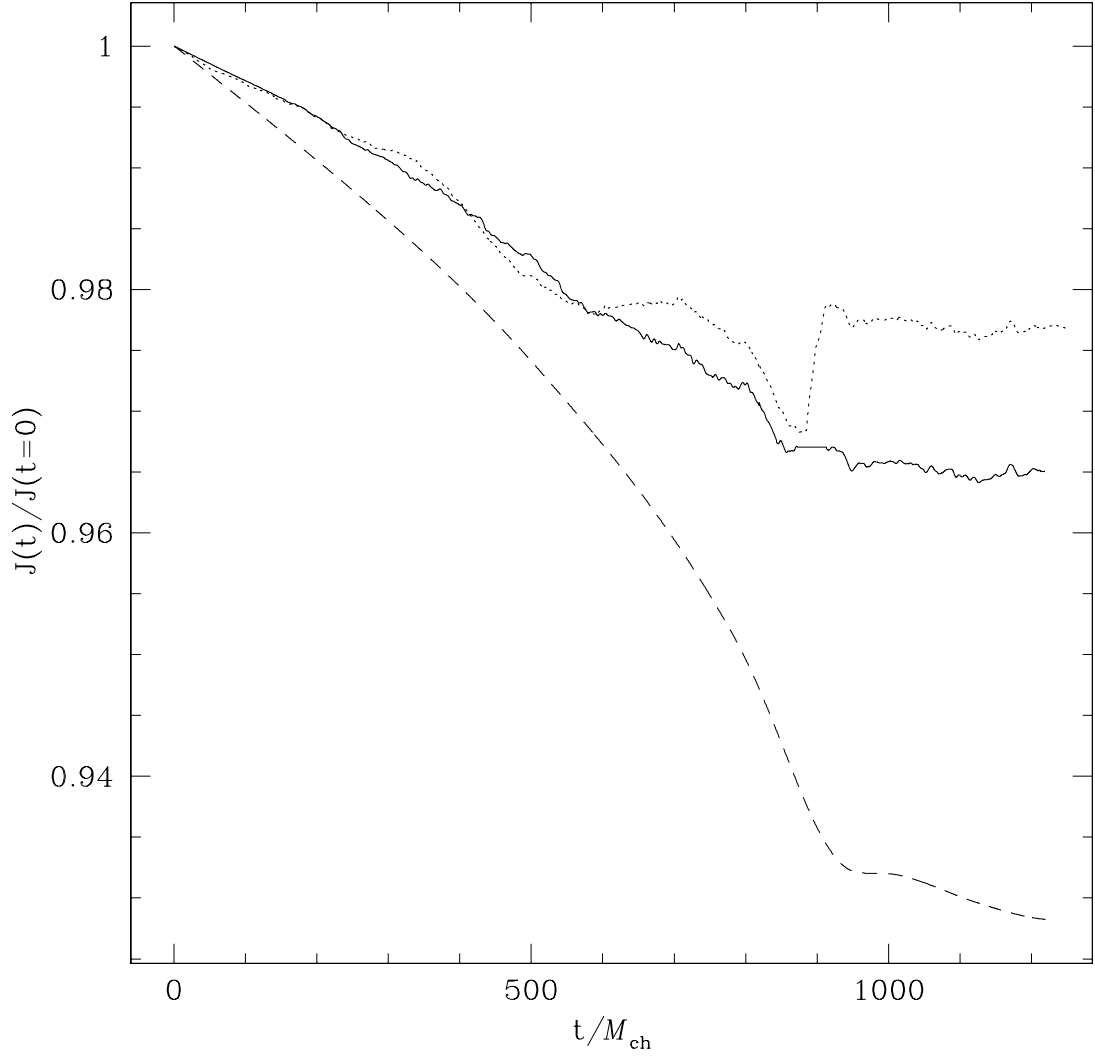}
\caption{Angular momentum loss for the binary in run RR1
measured using the  SPH integral,
  Eq.~\protect\ref{eq:j_sph} (dotted line), and the result after
  correcting for spurious angular momentum creation during the binary
  to single-object transition (solid line).  We
  see that the result yields an angular momentum loss approximately
  half that we would have predicted from the quadrupole formula,
  Eq.~\protect\ref{eq:j_quad} (dashed line), primarily because our
  estimate for the system's angular velocity used in the backreaction
  force is systematically lower
  than the GW signal would indicate.  The quadrupole result we
  derive shows that about $7\%$ of the system angular momentum is
  emitted in GWs, in line with previous relativistic estimates.} 
\label{fig:rr_jdiff}
\end{figure}

One possible cause for concern with our code is non-conservative
behavior caused by numerical errors that develop after
$t/{\cal M}_{ch}=500$.
In Fig.~\ref{fig:rr_jdiff}, we show
the change in the system's angular momentum over time.  The dotted
curve is the uncorrected result derived from our calculation, 
which shows two periods of angular
momentum generation, the first from $t/{\cal M}_{ch}=600-700$, 
and the second at
$t/{\cal M}_{ch}=875-900$.  
The former is associated with the numerical inaccuracies
discussed in Sec.~\ref{sec:stable} for run RR2.
All of our runs
show some spurious angular momentum generation and slowing of the binary
infall during this period.  The latter spike occurs immediately before
and after the transition from a binary to a single-star description.
Correcting for both of these spurious terms yields the solid line on
the plot, which still underestimates by a non-trivial amount the
angular momentum loss we would expect from the quadrupole formula,
\begin{equation}
(\dot{J}_k)_{GW}=0.4
  \epsilon_{ijk}\left<Q^{[2]}_{il}Q^{[3]}_{lk}\right>.
\label{eq:j_quad}
\end{equation}
Using the quadrupole formula on our results yields a total
angular momentum loss fraction which very nearly equals that found by
\cite{Oech,STU}, with approximately $7\%$ of the system's angular
momentum converted into GW emission.  The discrepancy between this
amount and what we derive from the SPH particle 
configuration, Eq.~\ref{eq:j_sph}, can be easily understood.  
First, we do not 
see strong angular 
momentum loss as the NS first make contact, since this is where we
push our field solver to its limit.  Second, our
method for estimating the instantaneous angular velocity,
Eq.~\ref{eq:omega_sph}, underestimates the proper 
value of $\omega$, yielding a
radiation reaction force, Eq,~\ref{eq:rrforce}, smaller than the
correct quadrupole value (which we can determine after the fact).
This error could be
decreased in magnitude by calculating the angular velocity from the
change of
position of the NS centers-of-mass in time, but such a prescription is
difficult to define consistently in the single-body regime.  Even defining
the orbital frequency in terms of the rate of change of the
quadrupole tensor, as done by \cite{Oech}, underestimates the
correct angular momentum loss rate by up to $40\%$ during the GW
emission peak.

\begin{figure}
\centering\includegraphics[width=6in]{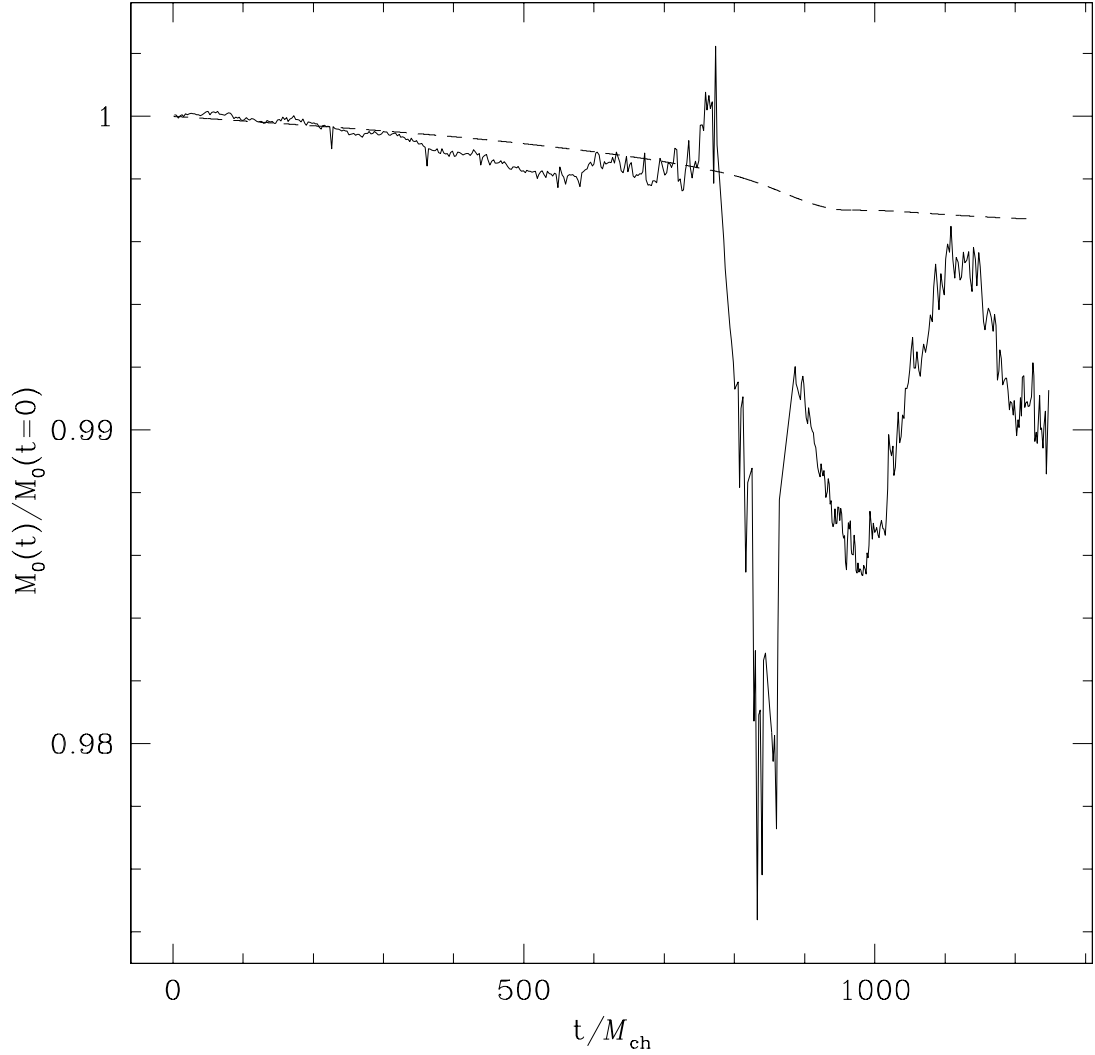}
\caption{The evolution of the total ADM mass for the system in run
  RR1, calculated
  using the SPH integral, Eq.~\protect\ref{eq:adm_sph} (solid line).  
Up until $t/{\cal M}_{ch}=600$,
  we see a slow decrease as energy is emitted in GWs, at
  approximately the rate predicted by the quadrupole formula,
  Eq.~\protect\ref{eq:dedtqe} (dashed line).  Beyond $t/{\cal M}_{ch}=600$ 
numerical inaccuracies
  lead to oscillations of total amplitude $\sim 2\%$.  From the
  quadrupole estimate, we find that $0.4\%$ of the system's energy is
  radiated away as GWs, in line with previous estimated from
  relativistic calculations. }
\label{fig:rr_adm}
\end{figure}

We see a similar pattern at work in the evolution of the system's ADM
mass, shown in Fig.~\ref{fig:rr_adm}, comparing the energy loss to
GWs from the quadrupole approximation formula (dashed line),
Eq.~\ref{eq:dedtqe}, 
to the value we find from SPH summation via Eq.~\ref{eq:adm_sph} (solid line). 
The quadrupole value agrees well with other calculations, which
typically find $\Delta M_{ADM}/M_{ADM}=0.004$.
Looking at our particle summation value, we see a slow decrease from
$t/{\cal M}_{ch}=0-500$, of magnitude $0.2-0.3\%$ of total value, in good agreement 
with other calculations and our own quadrupole estimate.  
From that point on, we see a small spurious
increase from the point the numerical errors begin to become
significant, followed by a sharp decrease of $\sim 2\%$ immediately
prior to our transition 
from binary to single-star descriptions. Once we have made the
transition, the ADM mass oscillates slightly, with an overall
peak-to-trough amplitude of $1\%$.  Thus, we conclude that while the
instantaneous value we measure from the particles directly is liable
to be off by up to $2\%$, we can reconstruct the proper energy loss
rate after the calculation is over.

\section{Conclusions}\label{sec:final}

We have developed and tested a new relativistic 3-d Lagrangian
hydrodynamics code, which should prove useful for studying a wide
variety of physical systems.  Here, as an initial investigation, we have
performed the first full evolutions of the coalescence and merger of
irrotational binary NS in
the CF approximation to GR.  Moreover, these calculations
represent the first numerical evolution of coalescing binary systems performed
with either a spectral methods field solver or the use of spherical
coordinates adapted to a binary environment. 

The code has been validated using several tests.
We can accurately reproduce static spherical stellar
configurations as well as the known solution for a collapsing
pressureless dust cloud.  In both cases, the CF approximation yields
the exact solution in GR, as it will for any spherical configuration.
We find that we can reproduce these known semi-analytic solutions to
high accuracy, up until the formation of an event horizon for the
collapsing dust clouds.

Our dynamical evolution calculations for 
quasi-equilibrium models at a number of
different binary separations indicates that we can successfully
integrate forward for several orbits, with typical errors in
conserved quantities of $\sim 1\%$.  In doing
so, we have demonstrated directly for the first time that the  
$M/R=0.14$, equal-mass sequence of TG is stable all the way to its innermost
configuration, at which point a cusp develops on the inner edge of
each NS.

Our dynamical calculation of a complete binary NS merger, including radiation
reaction effects, demonstrated that our
spherical coordinates,  
spectral method approach is robust enough to follow the 
system from the point just before the formation of a cusp 
through merger and 
the formation of a stable remnant.  Some errors were introduced
during this period into the globally conserved quantities such as the
ADM mass and system angular momentum, but we find that the field
values were computed consistently throughout, and that the global
dynamics was treated in a quantitatively accurate way.  

We find that the merger remnant formed in our calculation is
differentially rotating, with a transient quadrupole deformation.
This combination of effects produces a GW amplitude with a
modulated form, similar to what has been seen before in PN
calculations (FR3) and more recent full GR calculations of the
same model \cite{STU}.  We find that
the remnant is initially stable against gravitational collapse, as did
\cite{STU}, with the
supermassive NS (which has a 
baryonic mass essentially twice that of either NS in
isolation) supported by strong differential rotation.  We find that a
density maximum develops rather rapidly in the center of the merger
remnant, as has been seen in all other PN and CF calculations, 
but not that of \cite{STU}, whose full GR merger calculation yielded a
toroidal remnant.  We believe the difference results from their use of
a capturing scheme, whereas our runs were performed using an adiabatic 
treatment.

By combining our calculated GW signal with a relativistic
quasi-equilibrium inspiral precursor, we have generated the first
GW energy spectrum from a binary NS merger which is complete at all
sub-kHz frequencies and consistent throughout.  We find that the
energy spectrum deviates from the Newtonian power-law relation by more
than $50\%$ at frequencies $f_{GW}<$ 1 kHz (the ``break frequency''), 
in very good agreement
with the predictions of FGRT.  There are distinct peaks in the
power spectrum corresponding to the phases of maximum GW luminosity
and merger remnant ringdown, but at levels significantly below the
point-mass power-law value.

In our future work on binary NS systems, we hope to address a number
of topics, many of which deserve much more careful study.  Based on
the excellent 
agreement between our calculated GW energy spectrum and that based purely on
equilibrium sequence data, we hope to do a broad phase space survey to
determine the dependence of the ``break frequency'' on
both the NS EOS and the system's mass ratio.
Beyond this parameter study of NS-NS mergers, we also plan to investigate
in detail the formation process for the merger remnant, to determine
the conditions which may lead to the formation of a quasi-toroidal merger
remnant. This will necessarily involve the use of a relativistic
artificial viscosity scheme to treat shocks.  
The density profile of the merger remnant is likely to
influence the final fate of the system, and may prove crucial for
determining the coincidence properties of GW emissions and
short-period GRBs, should they result from compact object binary
mergers, as has been widely suggested.

\begin{acknowledgments}
This work was supported by NSF grants PHY-0133425 and
PHY-0245028 to Northwestern University.  We thank our anonymous
referee for numerous helpful suggestions.
\end{acknowledgments}

\bibliography{paper5bib}

\newpage
\begin{table}
\caption{A comparison of our notation for various relativistic
quantities to previous works using the CF formalism:
\protect\cite{GGTMB,Oech,SBS,WMM}.  For those cases where no unique
terminology was defined, we give the simplest equivalent algebraic form.
We also list the equation in this paper where the quantity is defined}
\begin{ruledtabular}\begin{tabular}{l|ccccc|c}
Quantity & Here & Gourgoulhon & Oechslin & Shibata & Wilson & See Equation\\
\colrule\colrule
Lapse & $N$ & $N$ & $\alpha$ & $\alpha$ & $\alpha$ & \protect\ref{eq:metric}\\
Shift & $N_i$ & $N_i$ & $-\beta_i$ & $-\beta_i$ & $-\beta_i$ &
\protect\ref{eq:metric}\\ 
Conf. Fact. & $A$ & $A$ & $\psi^2$ & $\psi^2$ & $\phi^2$ &
\protect\ref{eq:metric}\\ 
Rest dens. & $\rho_*$ & $\Gamma_n A^3 \rho$ & $\rho_*$ & $\rho_*$ &
$D\phi^6$ & \protect\ref{eq:rhostar} \\
Lorentz Fact. & $\gamma_n$ & $\Gamma_n$ & $\alpha u^0$ & $\alpha u^0$
& $W$ & \protect\ref{eq:lorentz}\\
Velocity & $v^i$ & $NU^i+N^i$ & $v^i$ & $v^i$ & $V^i$ & \protect\ref{eq:uv}\\
Spec. Momentum & $\tilde{u}_i$ & $w_i$ & $\tilde{u}_i$ & $\tilde{u}_i$
& $S_i/(D\phi^6)$ & \protect\ref{eq:utilde}\\
Enthalpy & $h$ & $h$ & $w$ & $1+\Gamma \epsilon$ & $h$ &
\protect\ref{eq:enthalpy} \\
\end{tabular}\end{ruledtabular}
\label{table:not}
\end{table}

\begin{table}
\caption{A summary of the runs presented in this paper.}
\begin{tabular}{c|l|c}
\hline\hline
Run & Description & See Sec. $\#$ \\ \hline
OV1 & Equilibrium OV model, $GM_0/Rc^2=0.126$ & \protect\ref{sec:ov} \\
OV2 & Non-equilibrium OV model, $R_0=1.1 R_{eq}$  & \protect\ref{sec:ov} \\
OV3 & Same as OV2, w/relaxation drag & \protect\ref{sec:ov}\\ \hline
DC5 & Collapsing dust cloud, $M_0=1$, $R_0=5$ & \protect\ref{sec:dust} \\
DC10 & Collapsing dust cloud,$M_0=1$, $R_0=10$ & \protect\ref{sec:dust} \\
DC50 & Collapsing dust cloud,$M_0=1$, $R_0=50$ & \protect\ref{sec:dust}
\\ \hline
QC1 & Quasi-circular binary orbit w/o Rad. Reac., $r_0/{\cal M}_{ch}=19.91$ &
\protect\ref{sec:circ} \\
QC2 & Quasi-circular binary orbit w/o Rad. Reac., $r_0/{\cal M}_{ch}=20.42$ &
\protect\ref{sec:circ} \\
QC3 & Quasi-circular binary orbit w/o Rad. Reac., $r_0/{\cal M}_{ch}=22.98$ &
\protect\ref{sec:circ} \\ \hline
RR1 & Full binary evolution w/rad. reac., $r_0/{\cal M}_{ch}=19.91$ &
\protect\ref{sec:unstable} \\
RR2 & Full binary evolution w/rad. reac., $r_0/{\cal M}_{ch}=22.98$ &
\protect\ref{sec:stable} \\ \hline \hline
\end{tabular}
\label{table:runs}
\end{table}

\newpage

\newpage

\clearpage

\clearpage

\clearpage

\clearpage

\clearpage

\clearpage

\clearpage

\clearpage

\clearpage

\clearpage

\clearpage

\clearpage

\clearpage

\clearpage

\clearpage

\clearpage

\clearpage

\clearpage

\clearpage

\end{document}